\documentclass[journal,draftclsnofoot,onecolumn,11pt]{IEEEtran}
\usepackage{latexsym,amssymb,amsmath,graphicx,epsfig,cite,bbm,float,subfig}
\usepackage{ifpdf}
\def\ninept{\def\baselinestretch{1.5}}
\ninept

\def\ninept{\def\baselinestretch{1.5}}
\ninept
\newcommand{\bx}{{\mathbf{x}}}
\newcommand{\bb}{{\mathbf{b}}}
\newcommand{\ba}{{\mathbf{a}}}

\newcommand{\bp}{{\mathbf{p}}}
\newcommand{\bmu}{{\boldsymbol{\mu}}}
\newcommand{\bsig}{{\boldsymbol{\sigma}}}
\newcommand{\bSig}{{\mathbf{\Sigma}}}
\newcommand{\bDelta}{{\mathbf{\Delta}}}
\newcommand{\bone}{{\mathbf{1}}}
\newcommand{\bq}{{\mathbf{q}}}
\newcommand{\bw}{{\mathbf{w}}}

\newcommand{\beps}{{\mathbf{\epsilon}}}

\newcommand{\cN}{{\mathcal{N}}}

\newcommand{\cB}{{\mathcal{B}}}
\newcommand{\cL}{{\mathcal{L}}}
\newcommand{\cE}{{\mathcal{E}}}

\newcommand{\be}{\begin{equation}}
\newcommand{\ee}{\end{equation}}
\newcommand{\bea}{\begin{eqnarray}}
\newcommand{\eea}{\end{eqnarray}}

\newcommand{\MB}{\left[\begin{array}}
\newcommand{\ME}{\end{array}\right]}

\renewcommand{\vec}[1]{\mbox{\boldmath${#1}$}}

\newcommand{\ei}{\end{itemize}}
\newcommand{\bi}{\begin{itemize}}

\newcommand{\Rp}{\mbox{$\mathbbm{R}^+$}}
\newcommand{\Rm}{\mbox{$\mathbbm{R}^+_m$}}

\newcommand{\vb}{\mbox{$\vec{b}$}}

\newtheorem{theorem}{\bf{Theorem}}[section]
\newcommand{\defi}{\stackrel{\bigtriangleup}{=}}

    \def\squarebox#1{\hbox to #1{\hfill\vbox to #1{\vfill}}}

\newcommand{\eps}{\mbox{$\epsilon$}}
\begin{document}
\title{Optimal Investment Under Transaction Costs: A Threshold
  Rebalanced Portfolio Approach} \author{Sait Tunc and Suleyman
  S. Kozat, {\em Senior Member}, IEEE\thanks{This work is supported in
    part by IBM Faculty Award and Outstanding Young Scientist Award
    Program from Turkish Academy of Sciences. Suleyman S. Kozat and
    Sait Tunc (\{skozat,saittunc\}@ku.edu.tr) are with the Competitive
    Signal Processing Laboratory at the Koc University, Istanbul,
    34450, tel: +902123381864, fax: +902123381548.}}  \maketitle

\begin{abstract}
We study optimal investment in a financial market having a finite
number of assets from a signal processing perspective.  We investigate
how an investor should distribute capital over these assets and when
he should reallocate the distribution of the funds over these assets
to maximize the cumulative wealth over any investment period. In
particular, we introduce a portfolio selection algorithm that
maximizes the expected cumulative wealth in i.i.d. two-asset
discrete-time markets where the market levies proportional transaction
costs in buying and selling stocks. We achieve this using ``threshold
rebalanced portfolios'', where trading occurs only if the portfolio
breaches certain thresholds. Under the assumption that the relative
price sequences have log-normal distribution from the Black-Scholes
model, we evaluate the expected wealth under proportional transaction
costs and find the threshold rebalanced portfolio that achieves the
maximal expected cumulative wealth over any investment period. Our
derivations can be readily extended to markets having more than two
stocks, where these extensions are pointed out in the paper. As
predicted from our derivations, we significantly improve the achieved
wealth over portfolio selection algorithms from the literature on
historical data sets.
\end{abstract}
\begin{center}
\bfseries EDICS Category: MLR-APPL, MLR-LEAR, SSP-APPL.
\end{center}

\section{Introduction}
\label{sec:introduction}

Recently financial applications attracted a growing interest from the
signal processing community since the recent global crises
demonstrated the importance of sound financial modeling and reliable
data processing \cite{speciss, speciss2}. Financial markets produce
vast amount of temporal data ranging from stock prices to interest
rates, which make them ideal mediums to apply signal processing
methods.  Furthermore, due to the integration of high performance,
low-latency computing recourses and the financial data collection
infrastructures, signal processing algorithms could be readily
leveraged with full potential in financial stock markets. This paper
particularly focuses on the portfolio selection problem, which is one
the most important financial applications and has already attracted
substantial interest from the signal processing community
\cite{Bean09, Torun11a, Bean10, Bean11,Torun11b}.

In particular, we study the investment problem in a financial market
having a finite number of assets. We concentrate on how an investor
should distribute capital over these assets and when he should
reallocate the distribution of the funds over those assets in time to
maximize the overall cumulative wealth.  In financial terms,
distributing ones capital over various assets is known as the
portfolio management problem and reallocation of this distribution by
buying and selling stocks is referred as the rebalancing of the given
portfolio \cite{investment}. Due to obvious reasons, the portfolio
management problem has been investigated in various different fields
from financial engineering \cite{markow}, machine learning to
information theory \cite{cover91}, with a significant room for
improvement as the recent financial crises demonstrated.  To this end,
we investigate the portfolio management problem in discrete-time
markets when the market levies proportional {\em transaction costs} in
trading while buying and selling stocks, which accurately models a
wide range of real life markets \cite{markow,investment}. In discrete
time markets, we have a finite number of assets and the reallocation
of wealth (or rebalancing of the capital) over these assets is only
allowed at discrete investment periods, where the investment period is
arbitrary, e.g., each second, minute or each day \cite{cover91,
  cover96}.  Under this framework, we introduce algorithms that
achieve the {\em maximal} expected cumulative wealth under
proportional transaction costs in i.i.d. discrete-time markets
extensively studied in the financial literature \cite{markow,
  investment}.  We further illustrate that our algorithms
significantly improve the achieved wealth over the well-known
algorithms in the literature on historical data sets under realistic
transaction costs, as anticipated from our derivations.  The precise
problem description including the market and transaction cost models
are provided in Section \ref{sec:trp}.

Determination of the optimum portfolio and the best portfolio
rebalancing strategy that maximize the wealth in discrete-time markets
with {\em no transaction fees} is heavily investigated in information
theory \cite{cover91, cover96}, machine learning
\cite{helmbold,ysinger, vovk98} and signal processing \cite{Kozat11,
  Kozat09, Kozat08, kozat_tree} fields. Assuming that the portfolio
rebalancings, i.e., adjustments by buying and selling stocks, require
no transaction fees and with some further mild assumptions on the
stock prices, the portfolio that achieves the maximum wealth is shown
to be a constant rebalanced portfolio (CRP) \cite{cover, cover96}.  A
CRP is a portfolio strategy where the distribution of funds over the
stocks are reallocated to a predetermined structure, also known as the
target portfolio, at the start of each investment period. CRPs
constitute a subclass of a more general portfolio rebalancing class,
the calendar rebalancing portfolios, where the portfolio vector is
rebalanced to a target vector on a periodic basis
\cite{investment}. Numerous studies are carried out to asymptotically
achieve the performance of the best CRP tuned to the individual
sequence of stock prices albeit either with different performance
bounds or different performance results on historical data sets
\cite{helmbold,cover96, vovk98}. CRPs under transaction costs are
further investigated in \cite{blum}, where a sequential algorithm
using a weighting similar to that introduced in \cite{cover}, is also
shown to be competitive under transaction costs, i.e., asymptotically
achieving the performance of the best CRP under transaction costs.
However, we emphasize that maintaining a CRP requires potentially
significant trading due to possible rebalancings at each investment
period \cite{Kozat11}. As shown in \cite{Kozat11}, even the
performance of the best CRP is severally affected by moderate
transaction fees rendering CRPs ineffective in real life stock
markets. Hence, it may not be enough to try to achieve the performance
of the best CRP if the cost of rebalancing outweighs that which could
be gained from rebalancing at every investment period.  Clearly, one
can potentially obtain significant gain in wealth by including
unavoidable transactions fees in the market model and perform
reallocation accordingly.

In these lines, the optimal portfolio selection problem under
transactions costs is extensively investigated for continuous-time
markets \cite{davis,taksar, morton, magill}, where growth optimal
policies that keep the portfolio in closed compact sets by trading
only when the portfolio hits the compact set-boundaries are
introduced. Naturally, the results for the continuous markets cannot
be straightforwardly extended to the discrete-time markets, where
continuous trading is not allowed. However, it has been shown in
\cite{iyengar02} that under certain mild assumptions on the sequence
of stock prices, similar no trade zone portfolios achieve the optimal
growth rate even for discrete-time markets under proportional
transaction costs. For markets having two stocks, i.e., two-asset
stock markets, these no trade zone portfolios correspond to threshold
portfolios, i.e., the no trade zone is defined by thresholds around
the target portfolio.  As an example, for a market with two stocks,
the portfolio is represented by a vector $\vec{b} = [b \; \; 1-b]^T$,
$b\in [0,1]$, assuming only long positions \cite{investment}, where
$b$ is the ratio of the capital invested in the first stock. For this
market, the no rebalancing region around a target portfolio $\vec{b} =
[b \; \; 1-b]^T$, $b\in [0,1]$, is given by a threshold $ \epsilon$,
$\min\{b,1-b\} \geq \epsilon \geq 0$, such that the corresponding
portfolio at any investment period is rebalanced to a desired vector
if the ratio of the wealth in the first stock breaches the interval
$(b-\epsilon,b+\epsilon)$.  In particular, unlike a calendar
rebalancing portfolio, e.g., a CRP, a threshold rebalanced portfolio
(TRP) rebalances by buying and selling stocks only when the portfolio
breaches the preset boundaries, or ``thresholds'', and otherwise does
not perform any rebalancing. Intuitively, by limiting the number of
rebalancings due to this non rebalancing regions, threshold portfolios
are able to avoid hefty transactions costs associated with excessive
trading unlike calendar portfolios. Although TRPs are shown to be
optimal in i.i.d. discrete-time two-asset markets (under certain
technical conditions) \cite{iyengar02}, finding the TRP that maximizes
the expected growth of wealth under proportional transaction costs is
not solved, except for basic scenarios \cite{iyengar02}, to the best
of our knowledge.

In this paper, we first evaluate the expected wealth achieved by a TRP
over any finite investment period given any target portfolio and
threshold for two-asset discrete-time stock markets subject to
proportional transaction fees. We emphasize that we study two-asset
market for notational simplicity and our derivations can be readily
extended to markets having more than two assets as pointed out in the
paper where needed.  We consider i.i.d. discrete-time markets
represented by the sequence of price relatives (defined as the ratio
of the opening price to the closing price of stocks), where the
sequence of price relatives follow log-normal distributions. Note that
the log-normal distribution is the assumed statistical model for price
relative vectors in the well-known Black-Scholes model
\cite{investment, markow} and this distribution is shown to accurately
model real life stock prices by many empirical studies
\cite{investment}. Under this setup, we provide an iterative relation
that efficiently and recursively calculates the expected wealth over
any period in {\em any} i.i.d. discrete time market. This iterative
relation is evaluated using a certain multivariate Gaussian integral
for the log-normal distribution. We then provide a randomized
algorithm to calculate the given integral and obtain the expected
growth. This expected growth is then optimized by a brute force method
to yield the optimal target portfolio and threshold to maximize the
expected wealth over any investment period. Furthermore, we also
provide a maximum-likelihood estimator to estimate the parameters of
the log-normal distribution from the sequence of price relative
vectors, which is incorporated into the algorithmic framework in
Simulations section since these parameters are naturally unknown in
real life markets.

Portfolio management problem is studied with transaction costs in
\cite{iyengar00} on the horse race setting, which is a special
discrete-time market where only one of the asset pays off and the
others pay nothing on each period. This basic framework is then
extended to general stock markets in \cite{iyengar02}, where threshold
portfolios are shown to be growth optimal for two-asset
markets. However, no algorithm, except for a special sampled Brownian
market, is provided to find the optimal target portfolio or threshold
in \cite{iyengar02}. To achieve the performance of the best TRP, a
sequential algorithm is introduced in \cite{iyengar_uni} that is shown
to asymptotically achieve the performance of the best TRP tuned to the
underlying sequence of price relatives. This algorithm uses a similar
weighting introduced in \cite{cover} to construct the universal
portfolio. We emphasize that the universal investment strategies,
e.g., \cite{iyengar_uni}, which are inspired by universal source
coding ideas, based on Bayesian type weighting, are heavily utilized
to construct sequential investment strategies \cite{Kozat11, Kozat08,
  Kozat09,kozat_tree,Bean09, Bean10, cover96,
  ysinger,vovk98}. Although these methods are shown to
``asymptotically'' achieve the performance of the best portfolio in
the competition class of portfolios, their non-asymptotic performance
is acceptable only if a sufficient number of candidate algorithms in
the competition class is overly successful \cite{Kozat11} to
circumvent the loss due to Bayesian type averaging.  Since these
algorithms are usually designed in a min-max (or universal) framework
and hedge against (or should even work for) the worst case sequence,
their average (or generic) performance may substantially suffer
\cite{borodin,barron,helmbold}. In our simulations, we show that our
introduced algorithm readily outperforms a wide class of universal
algorithms on the historical data sets, including
\cite{iyengar_uni}. Note that to reduce the negative effect of the
transaction costs in discrete time markets, semiconstant rebalanced
portfolio (SCRP) strategies have also been proposed and studied in
\cite{blum, helmbold, Kozat11}. Different than a CRP and similar to
the TRPs, an SCRP rebalances the portfolio only at the determined
periods instead of rebalancing at the start of each period. Since for
an SCRP algorithm rebalancing occurs less frequently than a CRP, using
an SCRP strategy may improve the performance over CRPs when
transaction fees are present. However, no formulation exists to find
the optimal rebalancing times for SCRPs to maximize the cumulative
wealth. Although there exist universal methods \cite{Kozat11, ysinger}
that achieve asymptotically the performance of the best SCRP tuned to
the underlying sequence of price relatives, these methods suffer in
realistic markets since they are tuned to the worst case scenario
\cite{Kozat11} as demonstrated in the Simulations section.

We begin with the detailed description of the market and the TRPs in
Section~\ref{sec:prob_desc}. We then calculate the expected wealth
using a TRP in an i.i.d. two-asset discrete-time market under
proportional transaction costs over any investment period in
Section~\ref{sec:trp}. We first provide an iterative relation to
recursively calculate the expected wealth growth. The terms in the
iterative algorithm are calculated using a certain form of
multivariate Gaussian integrals. We provide a randomized algorithm to
calculate these integrals in
Section~\ref{sec:multivariate-gaussian}. The maximum-likelihood
estimation of the parameters of the log-normal distribution is given
in Section~\ref{sec:mle_est}. The paper is then concluded with the
simulations of the iterative relation and the optimization of the
expected wealth growth with respect to the TRP parameters using the ML
estimator in Section~\ref{sec:sim}.

\section{Problem Description\label{sec:prob_desc}}
In this paper, all vectors are column vectors and represented by
lower-case bold letters. Consider a market with $m$ stocks and let
$\{\bx(t)\}_{t \geq 1}$ represent the sequence of price relative
vectors in this market, where $\bx(t) = [x_1(t),x_2(t), \ldots,
  x_m(t)]^T$ with $x_i(t) \in \Rp$ for $i \in \{1,2,\ldots,m\}$ such
that $x_i(t)$ represents the ratio of the closing price of the $i$th
stock for the $t$th trading period to that from the $(t-1)$th trading
period. At each investment period, say period $t$, $\bb(t)$ represents
the vector of portfolios such that $b_i(t)$ is the fraction of money
invested on the $i$th stock. We allow only long-trading such that
$\sum_{i=1}^m b_i(t)=1$ and $b_i(t) \geq 0$. After the price relative
vector $\bx(t)$ is revealed, we earn $\bb^T(t)\bx(t)$ at the period
$t$. Assuming we started investing using 1 dollars, at the end of $n$
periods, the wealth growth in a market with no transaction costs is
given by
\begin{align}
S(n) & = \prod_{t=1}^n \bb^T(t) \bx(t).
\label{eq:gain1}
\end{align}
If we use a CRP  \cite{cover91}, then we earn
\[
 \prod_{t=1}^n \bb^T \bx(t),
\]
at the end of $n$ periods ignoring the transaction costs. This method
is called ``constant rebalancing'' since at the start of each
investment period $t$, the portfolio vector $\bb(t) = [b_1(t), b_2(t),
  \ldots, b_m(t)]$ is adjusted, or rebalanced, to a predetermined
constant portfolio vector, say, $\bb = [b_1, b_2, \ldots, b_m]$ where
$\sum_{i=1}^m b_i = 1$. As an example, at the start of each investment
period $t$, since we invested using $\vb$ at the investment period $t-1$
and observed $x(t-1)$, the current portfolio vector, say
$\bb_{\mathrm{old}}(t)$,
\[
 \bb_{\mathrm{old}}(t) \defi \left[ \frac{b_1 x_1(t-1)}{ \sum_{i=1}^m b_i x_i(t-1) },\ldots, \frac{b_m x_m(t-1)}{\sum_{i=1}^m b_i x_i(t-1)} \right]^T,
\]
should be adjusted back to $\bb$. If we assume a symmetric
proportional transaction cost with cost ratio $c$, $0 \leq c \leq 1$,
for both selling and buying, then we need to spend approximately
$\sum_{i=1}^{m}b_{i,\mathrm{old}}(t)S(t) |b_{i,\mathrm{old}}(t)-b_i|
c$ dollars for rebalancing. Note that if the transaction costs are not
symmetric, the analysis follows by assuming
$c=c_{\mathrm{sell}}+c_{\mathrm{buy}}$ by \cite{blum}, where
$c_{\mathrm{sell}}$ and $c_{\mathrm{buy}}$ are the proportional
transaction costs in selling and buying, respectively. Since a CRP
should be rebalanced back to its initial value at the start of each
investment period, a transaction fee proportional to the wealth growth
up to the current period, i.e., $S(t)$, is required for each period
$t$. Hence, constantly rebalancing at each time $t$ may be unappealing
for large $c$.

To avoid such frequent rebalancing, we use TRPs, where we denote a TRP
with a target vector $\bb$ and a threshold $\beps$ (with certain
abuse of notation) as ``TRP with $(\bb,\beps)$''. For a sequence of
price relatives vectors $\bx^n \defi [\bx(1),\bx(2),\ldots,\bx(n)]$
with $\bx \in \Rm$, a TRP with $(\bb,\beps)$ rebalances the portfolio
to $\bb$ at the first time $\tau$ satisfying
\begin{align}
	\frac{b_j\prod_{t=1}^{\tau} x_j(t)}{\sum_{k=1}^m b_k\prod_{t=1}^{\tau} x_k(t)} \notin [b_j-\eps_j,b_j+\eps_j]
\end{align}
for any $j \in \{1,2,\ldots,m\}$, thresholds
$\epsilon_j$, and does not rebalance otherwise, i.e., while the
portfolio vector stays in the no rebalancing region. Starting from
the first period of a no rebalancing region, i.e., where the
portfolio is rebalanced to the target portfolio $\bb$, say $t=1$ for this
example, the wealth gained during any no rebalancing region is given
by
\begin{align}
	W(\bx^n | \bb^n \in \cE_{n}^{\mathrm{nc}}) = \sum_{k=1}^m b_k \prod_{t=1}^{n} x_k(t),
	\label{eq:gen_wealth}
\end{align}
where $\bb^n=[\bb(1),\bb(2),\ldots,\bb(n)]$ with $\bb(t)$ is the
portfolio at period $t$ and $\cE_{n}^{\mathrm{nc}}$ is the length $n$
no rebalancing region defined as
\begin{align}
	\cE_{n}^{\mathrm{nc}}= \{ \bb^n \,\, | \,\, \bb(1) = \bb, b_j(t) \in (b_j - \eps_j, b_j + \eps_j),  j \in \{1,2,\ldots,m \}, t \in \{1,2,\ldots,n\} \}.
	\label{eq:no-rebalancing}
\end{align}
A TRP pays a transaction fee when the portfolio vector leaves the predefined no rebalancing region, i.e., goes out of the 
no rebalancing region $\cE_{n}^{\mathrm{nc}}$, 
and rebalanced back to its target portfolio vector $\bb$. 
Since the TRP may avoid constant rebalancing, it may avoid excessive transaction fees while securing the portfolio to stay close the target 
portfolio $\bb$, when we have heavy transaction costs in the market.

\begin{figure}
\centering
\includegraphics[scale=0.50]{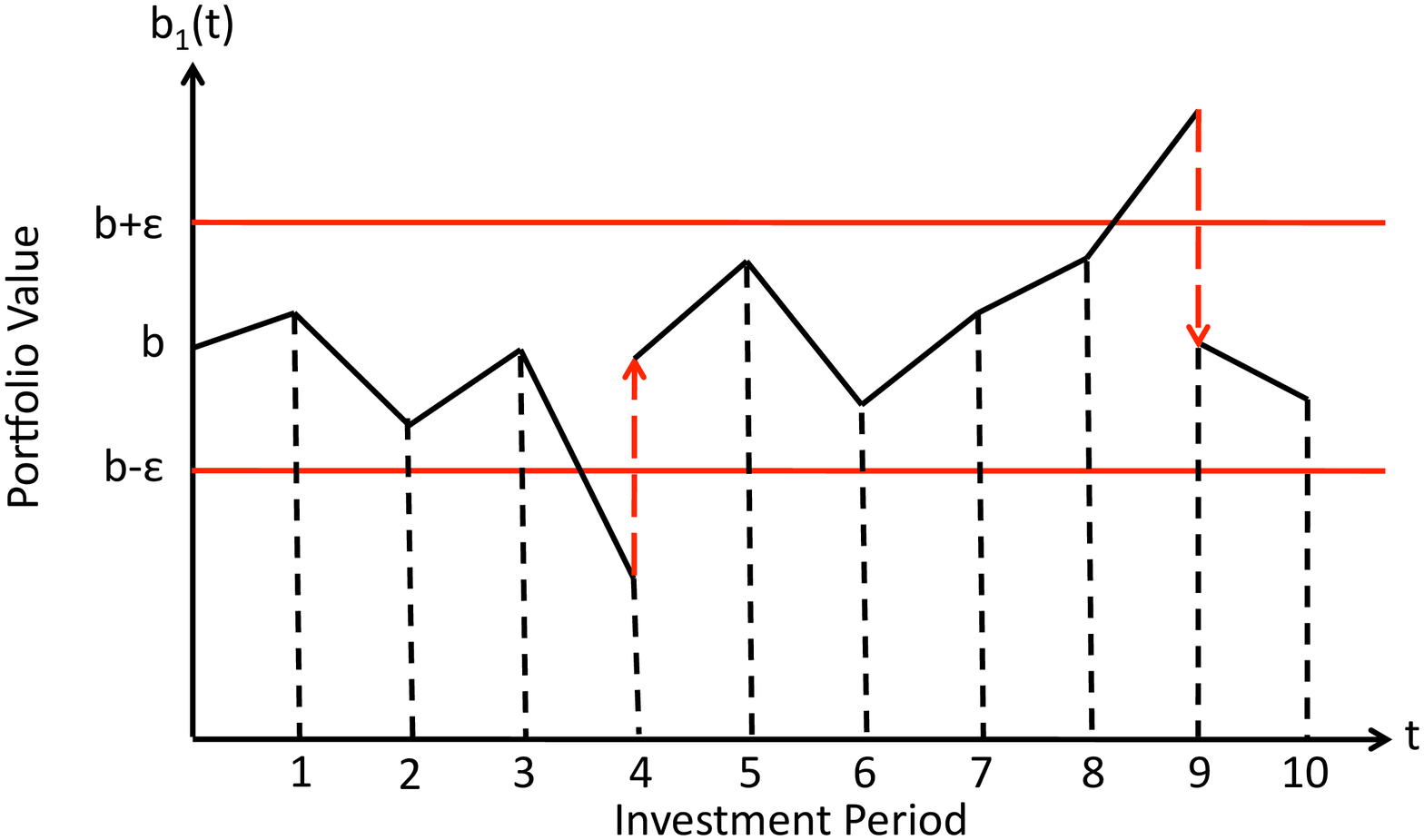}
\caption{\small A sample scenario for threshold rebalanced portfolios.}
\label{fig:sampleportfolio}
\end{figure}

For notational clarity, in the remaining of the paper, we assume that
the number of stocks in the market is equal to 2, i.e., $m=2$. Note
that our results can be readily extended to the case when $m > 2$. We
point out the necessary modifications to extend our derivations to
the case $m>2$.  Then, the threshold rebalanced portfolios are described
as follows. 

Given a TRP with target portfolio $\bb = [b,1-b]^T$ with $b \in
[0,1]$ and a threshold  $\beps$, the no
rebalancing region of a TRP with $(\bb,\beps)$ is represented by
$(b-\eps,b+\eps)$. Given a TRP with
$(b-\eps,b+\eps)$, we only rebalance if the portfolio leaves this region,
which can be found using only the first entry of the portfolio
(since there are two stocks), i.e., if $b_{1, \mathrm{old}}(t)
\notin (b-\eps, b+\eps)$.  In this case, we rebalance
$b_{1,\mathrm{old}} (t)$ to $b$. Fig.~\ref{fig:sampleportfolio}
represents a sample TRP in a discrete-time two-asset market and when
the portfolio is rebalanced back to its initial value if it leaves the no
rebalancing interval.

Before our derivations, we emphasize that the performance of a TRP is
clearly effected by the threshold and the target portfolio. As an
example, choosing a small threshold $\eps$, i.e., a low threshold, may
cause frequent rebalancing, hence one can expect to pay more
transaction fees as a result. However, choosing a small $\eps$ secures
the TRP to stay close to the target portfolio $\vb$. Choosing a larger
threshold $\eps$, i.e., a high threshold, avoids frequent rebalancing
and degrades the excessive transaction fees. Nevertheless, the
portfolio may drift to risky values that are distant from the target
portfolio $\vb$ under large threshold. Furthermore, we emphasize that
proportional transaction cost $c$ is a key factor in determining the
$\eps$. Under mild stochastic assumptions it has been shown in
\cite{cover,cover96} that in a market with no transaction costs, CRPs
achieve the maximum possible wealth. Therefore in a market with no
transaction costs, i.e., $c=0$, the maximum wealth can be achieved
when we choose a zero threshold, i.e., $\eps = 0$ and a target
portfolio $b^* = \underset{b}{\arg \max} \, E[\log(b x_1 + (1-b)
  x_2)]$, where $x_1$ and $x_2$ represent the price relatives of
two-asset market \cite{cover}. On the other hand, in a market with
high transaction costs, choosing a high threshold, i.e., a large 
$\eps$, eliminates the unappealing effect of transaction costs. For
instance, for the extreme case where the transaction cost is infinite,
i.e., $c=\infty$, the best TRP should either have
$\eps=1$ or $b \in \{ 0,1\}$ to ensure that no rebalancing occurs.

In this paper, we assume that the price relative vectors have a
log-normal distribution following the well-known Black-Scholes model
\cite{investment}. This distribution that is extensively used in
financial literature is shown to accurately model empirical price
relative vectors \cite{Bodie}. Hence, we assume that $\bx(t) =
[x_1(t), x_2(t)]^T$ has an i.i.d.  log-normal distribution with mean
$\bmu = [\mu_1 , \mu_2]$ and standard deviation $\bsig = [\sigma_1,
  \sigma_2]$, respectively, i.e., $\bx(t) \,\sim \, \ln \cN
(\bmu,\bsig^2)$. Here, we first optimize the wealth achieved
by a TRP for the discrete-time market, where the distributions of the
price relatives are known. We then provide a ML estimator for these
parameters to cover the case where the means and variances are
unknown. The ML estimator is incorporated in the algorithmic framework
in the Simulations section since the corresponding parameters are
unknown in real life markets.  The details of the maximum-likelihood
estimation are given in Section~\ref{sec:mle_est}.

\section{Threshold Rebalanced Portfolios \label{sec:trp}}
In this section, we analyze the TRPs in a discrete-time market with
proportional transaction costs as defined in
Section~\ref{sec:prob_desc}. We first introduce an iterative relation,
as a theorem, to recursively evaluate the expected achieved wealth of
a TRP over any investment period.  The terms in this iterative
equation are calculated using a certain form of multivariate Gaussian
integrals. We provide a randomized algorithm to calculate these
integrals. We then use the given iterative equation to find the
optimal $\eps$ and $b$ that maximize the expected wealth over any
investment period.

\subsection{An Iterative Relation to Calculate the Expected Wealth \label{sec:iteration}}
\begin{figure}
\centering
\includegraphics[scale=0.50]{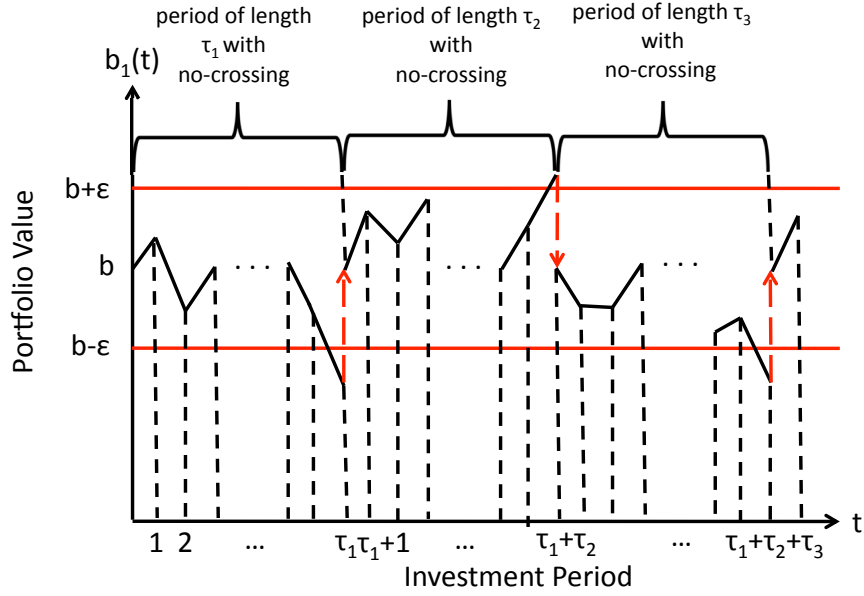}
\caption{\small No-crossing intervals of threshold rebalanced portfolios.}
\label{fig:nocrossing}
\end{figure}

In this section, we introduce an iterative equation to evaluate the
expected cumulative wealth of a TRP with $(b-\eps,b+\eps)$ over any
period $n$, i.e., $E[S(n)]$.  As seen in Fig.~\ref{fig:nocrossing},
for a TRP with $(b-\eps,b+\eps)$, any investment scenario can be
decomposed as the union of consecutive no-crossing blocks such that
each rebalancing instant, to the initial $\bb$, signifies the end of a
block. Hence, based on this observation, the expected gain of a TRP
between any consecutive crossings, i.e., the gain during the no
rebalancing regions, directly determines the overall expected wealth
growth. Hence, we first calculate the conditional expected gain of a
TRP over no rebalancing regions and then introduce the iterative
relation based on these derivations.

For a TRP with $(b-\eps,b+\eps)$, we call a no rebalancing region of
length $n$ as ``period $n$ with no-crossing'' such that the TRP with
the initial and target portfolio $\bb = [b,1-b]$ stays in the
$(b-\eps, b+\eps)$ interval for $n-1$ consecutive investment periods
and crosses one of the thresholds at the $n$th period. We next
calculate the expected gain of a TRP over any no-crossing period as
follows.

The wealth growth of a TRP with $(b-\eps,b+\eps)$ for a period $\tau$ with no-crossing can be written as \footnote{This is the special case of \eqref{eq:gen_wealth} for $m=2$.}
\begin{equation}
S_{\mathrm{nc}}(\tau) \defi b \prod_{t=1}^\tau [x_1(t)] + (1-b)  \prod_{t=1}^\tau [x_2(t)],
\label{eq:no-cross-gain}
\end{equation}
without the transaction cost that arises at the last period. To find
the total achieved wealth for a period $\tau$ with no-crossing, we
need to subtract the transaction fees from
\eqref{eq:no-cross-gain}. If portfolio $b_1(t)$ crosses the threshold
at the investment period $t=\tau$, then we need to rebalance it back
to $b$, i.e., $b_1(t)=b$ and  pay
\begin{equation}
S_{\mathrm{nc}}(\tau) c \left| \frac{ b \prod_{t=1}^{\tau} (x_1(t)) }{ b \prod_{t=1}^{\tau} (x_1(t)) + (1-b) \prod_{t=1}^{\tau} (x_2(t)) }-b \right |,
\end{equation}
where $c$ represents the symmetrical commission cost, to rebalance two stocks, i.e., $b_{1,\mathrm{old}}(\tau+1)$ to $b$,
and $b_{2,\mathrm{old}}(\tau+1) = 1 -b_{1,\mathrm{old}}(\tau+1)$ to $1-b$. Hence, the net overall gain for a period $\tau$ with no-crossing becomes
\begin{align}
&S(\tau)=S_{\mathrm{nc}}(\tau)- S_{\mathrm{nc}}(\tau) c \left| \frac{ b \prod_{t=1}^{\tau} (x_1(t)) }{ b \prod_{t=1}^{\tau} (x_1(t)) + (1-b) \prod_{t=1}^{\tau} (x_2(t)) }-b \right | \nonumber \\
& = b \prod_{t=1}^{\tau} [x_1(t)] + (1-b) \prod_{t=1}^{\tau} [x_2(t)] - c(b - b^2) \left| \prod_{t=1}^{\tau} [x_1(t)] - \prod_{t=1}^{\tau} [x_2(t)] \right| \nonumber \\
& = \zeta_1 \prod_{t=1}^{\tau} [x_1(t)] + \zeta_2 \prod_{t=1}^{\tau} [x_2(t)] \label{eq:nocrossgain},
\end{align}
where $\zeta_1 \defi b - 2 c (b - b^2)$ and $\zeta_2 \defi 1- b + 2 c
(b - b^2)$ for $b+\eps$ hitting and $\zeta_1 \defi b + 2 c (b - b^2)$
and $\zeta_2 \defi 1- b - 2 c (b - b^2)$ for $b-\eps$ hitting. Thus,
the conditional expected gain of a TRP conditioned on that the
portfolio stays in a no rebalancing region until the last period of
the region can be found by calculating the expected value of
\eqref{eq:nocrossgain}. Since, we now have the conditional expected gains,
we next introduce an iterative relation to find the expected wealth
growth of a TRP with $(b-\eps,b+\eps)$ for period $n$, $E[S(n)]$, by
using the expected gains of no-crossing periods as shown in
Fig.~\ref{fig:nocrossing}.

In order to calculate the expected wealth $E[S(n)]$ iteratively, let
us first define the variable $R(\tau)$, which is the expected
cumulative gain of all possible portfolios that hit
any of the thresholds first time at the $\tau$th period, i.e.,
\begin{align}
	R(\tau) = E \left[ S(\tau) \, {\Big |} \, \bb^{\tau} \in \cE_{\tau}^{\mathrm{fc}} \right], \label{eq:R}
\end{align}
where $\cE_{\tau}^{\mathrm{fc}}$ denotes the set of all possible portfolios with initial portfolio $b$ and that stay in the no rebalancing region for $\tau-1$ consecutive periods and hits one of the $b-\eps$ or $b+\eps$ boundary at the $\tau$th period, i.e.,
\begin{align}
	 \cE_{\tau}^{\mathrm{fc}} \defi \{ \bb^{\tau} \in \cB_{\tau}(b,\eps) \, | \, b(1) = b, b(i) \in (b-\eps,b+\eps) \, \forall i \in \{ 2,\ldots, \tau -1 \}, b(\tau) \notin (b-\eps, b+\eps) \}. \label{eq:fc}
\end{align}
Here, $\cB_\tau(b,\eps)$ is defined as the set of all possible
threshold rebalanced portfolios with initial and target portfolio $b$ and a no
rebalancing interval $(b-\eps,b+\eps)$. Similarly we define the
variable $T(\tau)$, which is the expected growth of all possible
portfolios of length $\tau$ with no threshold crossings, i.e.,
\begin{align}
	T(\tau) = E \left[ S(\tau) \,{\Big |}\, \bb^{\tau} \in \cE_{\tau}^{\mathrm{nc}} \right], \label{eq:T}
\end{align}
where $\cE_{\tau}^{\mathrm{nc}}$ denotes the set of portfolios with initial portfolio $b$ and that stay in the no rebalancing region for $\tau$ consecutive periods, i.e.,\footnote{This is the special case of the definition in \eqref{eq:no-rebalancing} for $m=2$.}
\begin{align}
	 \cE_{\tau}^{\mathrm{nc}} \defi \{ \bb^{\tau} \in \cB_{\tau}(b,\eps) \,| \, b(1) = b, b(i) \in [b-\eps,b+\eps] \, \forall i \in \{2,\ldots, \tau \} \}. \label{eq:nc}
\end{align}

Given the variables $R(\tau)$ and $T(\tau)$, we next introduce a
theorem that iteratively calculates the expected wealth growth of a
TRP over any period $n$. Hence, to calculate the expected achieved
wealth, it is sufficient to calculate $R(\tau)$, $T(\tau)$, threshold
crossing probabilities $P \left( \bb^n \in
\cE_{n}^{\mathrm{fc}}\right)$ and $P\left( \bb^n \in
\cE_{n}^{\mathrm{nc}}\right)$, which are explicitly evaluated in the
next section. \\

\noindent
\begin{theorem}
The expected wealth growth of a TRP $(b-\eps, b+\eps)$, i.e., $E[S(n)]$, over any i.i.d. sequence of price relative vectors $\bx^n = [\bx(1), \bx(2),\ldots,\bx(n)]$,  satisfies
\begin{align}
E[S(n)] = \sum_{i=1}^{n} P(\cE_{i}^{\mathrm{fc}}) R(i) E[S(n-i)] + P(\cE_{n}^{\mathrm{nc}}) T(n),
\label{eq:iterative-gain}
\end{align}
where we define $S_0 = 1$, $R(n)$ in \eqref{eq:R}, $T(n)$ in
\eqref{eq:T}, $\cE_{i}^{\mathrm{fc}}$ in \eqref{eq:nc} and
$\cE_{n}^{\mathrm{nc}}$ in \eqref{eq:fc}. \\
\label{thm:iterative_wealth}
\end{theorem} 

We emphasize that by Theorem~\ref{thm:iterative_wealth}, we can
recursively calculate the expected growth of any TRP over any
i.i.d. discrete-time market under proportional transaction
costs. Theorem~\ref{thm:iterative_wealth} holds for i.i.d. markets
having either  $m=2$ or $m >2$ provided that the corresponding terms
in \eqref{eq:iterative-gain} can be calculated. \\

\begin{proof}
By using the law of total expectation \cite{woods}, $E[S(n)]$ can be
written as
\begin{align}
E[S(n)] = \int_{\bb^n \in \cB_n(b,\eps)} E[S(n) | \bb^n] P(\bb^n) \mathrm{d}\bb^n,
\label{eq:iterative1}
\end{align}
where $\cB_n(b,\eps)$ is defined as the set of all possible TRPs with
the initial and target portfolio $b$ and threshold $\eps$. To obtain
\eqref{eq:iterative-gain}, we consider all possible portfolios as a
union of $n+1$ disjoint sets: (1) the portfolios which cross one of
the thresholds first time at the $1$st period; (2) the portfolios
which cross one of the thresholds first time at the $2$nd period; 
and continuing in this manner, (3) the portfolios which cross one of the
thresholds first time at the $n$th period; and finally (4) the portfolios
which do not cross the thresholds for $n$ consecutive
periods. Clearly these market portfolio sets are disjoint and their
union provides all possible portfolio paths. Hence \eqref{eq:iterative1}
can also be written as
\begin{align}
E[S(n)] &= \sum_{i=1}^n \,\, \int_{\bb_1^i \in \cE_{i}^{\mathrm{fc}} ,\bb_{i+1}^n \in \cB{n-i}(b,\eps) } E[S(n) | \bb_1^i \in \cE_{i}^{\mathrm{fc}} ,\bb_{i+1}^n \in \cB_{n-i}(b,\eps)] P(\bb_1^i \in \cE_{i}^{\mathrm{fc}} ,\bb_{i+1}^n \in \cB_{n-i}(b,\eps)) \mathrm{d}\bb^n \nonumber \\
&+ \int_{\bb^n \in \cE_{n}^{\mathrm{nc}}} E[S(n) | \bb^n \in \cE_{n}^{\mathrm{nc}}] P(\bb^n \in \cE_{n}^{\mathrm{nc}}) \mathrm{d}\bb^n,
\label{eq:iterative2}
\end{align}
where $\bb_i^j \defi [b(i),b(i+1),\ldots,b(j)]$. To continue with our
derivations, we define $S_{i \rightarrow j}$ as the wealth
growth from the period $i$ to period $j$, i.e., $S_{i \rightarrow j}
\defi \frac{S(j)}{S(i)}$. Assume that in the period $\tau$, the
portfolio crosses one of the thresholds and a rebalancing occurs. In
that case, regardless of the portfolios before the period
$\tau$, the portfolio is rebalanced back to its initial value in the
$\tau$th period, i.e., to $[b,1-b]^T$. Since the price relative
vectors are independent over time, we can conclude that the portfolios
before the period $\tau$ are independent from the portfolios after the
period $\tau$, i.e., $b(\tau)=b$ and every portfolio $b(i)$ for $i \in \{
1,2,\ldots,\tau-1\}$ are independent from the portfolios $b(j)$ for $j
\in \{\tau+1,\tau+2,\ldots,n\}$. Hence, the investment period
where the portfolio path crosses one of the thresholds, i.e., $\tau$,
divides the whole investment block into uncorrelated blocks in terms
of price relative vectors and portfolios. Thus, the wealth growth
acquired up to the period $\tau$, $S_{1 \rightarrow \tau}$, is
uncorrelated to the wealth growth acquired after that period, i.e.,
$S_{\tau+1 \rightarrow n}$. Hence, if we assume that a threshold
crossing occurs at the period $\tau$, then we have
\begin{align}
	E\left[S(n) | \bb_1^{\tau} \in \cE_{\tau}^{\mathrm{fc}} ,\bb_{\tau+1}^n \in \cB_{n-\tau}(b,\eps)\right] \,& =\, E\left[S_{1 \rightarrow \tau} S_{\tau+1 \rightarrow n} | \bb_1^{\tau} \in \cE_{\tau}^{\mathrm{fc}} ,\bb_{\tau+1}^n \in \cB_{n-\tau}(b,\eps) \right] \,\nonumber \\
	&=\, E\left[S_{1 \rightarrow \tau} | \bb_1^i \in \cE_{i}^{\mathrm{fc}} \right] \, E\left[S_{\tau+1 \rightarrow n} | \bb_{i+1}^n \in \cB_{n-i}(b,\eps)\right]. \label{eq:iterative2a}
\end{align} 
Applying \eqref{eq:iterative2a} to \eqref{eq:iterative2}, we get
\begin{align}
E\left[S(n)\right] &= \sum_{i=1}^n \,\, \int_{\bb_1^i \in \cE_{i}^{\mathrm{fc}} ,\bb_{i+1}^n \in \cB_{n-i}(b,\eps) } E\left[S_{1 \rightarrow i} | \bb_1^i \in \cE_{i}^{\mathrm{fc}}\right] E\left[S_{i+1 \rightarrow n} | b(i) = b, \bb_{i+1}^n \in \cB_{n-i}(b,\eps)\right] P\left(\bb_1^i \in \cE_{i}^{\mathrm{fc}}\right) \nonumber \\ 
&\times P\left(\bb_{i+1}^n \in \cB_{n-i}(b,\eps)\right) \mathrm{d}\bb^n + \int_{\bb^n \in \cE_{n}^{\mathrm{nc}}} E\left[S(n) | \bb^n \in \cE_{n}^{\mathrm{nc}}\right] P\left(\bb^n \in \cE_{n}^{\mathrm{nc}}\right) \mathrm{d}\bb^n.
\label{eq:iterative3}
\end{align}
Since the integral in \eqref{eq:iterative3} can be decomposed into two
disjoint integrals, \eqref{eq:iterative2} yields
\begin{align}
E[S(n)] &= \sum_{i=1}^n \,\, \int_{\bb_1^i \in \cE_{i}^{\mathrm{fc}}} E[S_{1 \rightarrow i} | \bb_1^i \in \cE_{i}^{\mathrm{fc}}] P(\bb_1^i \in \cE_{i}^{\mathrm{fc}}) \mathrm{d}\bb_1^i 
 \int_{\bb_{i+1}^n \in \cB_{n-i}(b,\eps)}  E[S_{i+1 \rightarrow n} | b(i) = b, \bb_{i+1}^n \in \cB_{n-i}(b,\eps)] \nonumber \\
& \times P(\bb_{i+1}^n \in \cB_{n-i}(b,\eps)) \mathrm{d}\bb_{i+1}^n + \int_{\bb^n \in \cE_{n}^{\mathrm{nc}}} E[S(n) | \bb^n \in \cE_{n}^{\mathrm{nc}}] P(\bb^n \in \cE_{n}^{\mathrm{nc}}) \mathrm{d}\bb^n .
\label{eq:iterative4}
\end{align}
We next write \eqref{eq:iterative4} as a recursive equation. 

To accomplish this, we first note that \\
\noindent
(i)  $R(i)$ is defined as the expected gain of TRPs with length $i$,
which crosses one of the thresholds first time at the $i$-th period and it
follows that
\begin{align}
R(i) & = E \left[ S(\tau) \, {\Big |} \, \bb^{i} \in \cE_{i}^{\mathrm{fc}} \right] \\
 & = \frac{1}{P(\cE_{i}^{\mathrm{fc}})} \int_{\bb_1^i \in \cE_{i}^{\mathrm{fc}}} E[S_{1 \rightarrow i} | \bb_1^i \in \cE_{i}^{\mathrm{fc}}] P(\bb_1^i \in \cE_{i}^{\mathrm{fc}}) \mathrm{d}\bb_1^i,
\label{eq:iterative5}
\end{align}
where we write $P(\cE_{i}^{\mathrm{fc}})$ instead of $P(\bb_1^i \in \cE_{i}^{\mathrm{fc}})$. \\ 
\noindent
(ii) Then, as the second term,  $T(n)$ is defined as the expected gain of TRPs of length
$n$, which does not cross one of the thresholds for $n$ consecutive
periods. This yields
\begin{align}
T(n) & = E \left[ S(n) \,{\Big |}\, \bb^{n} \in \cE_{n}^{\mathrm{nc}} \right]\\
 & = \frac{1}{P(\cE_{n}^{\mathrm{nc}})} \int_{\bb^n \in \cE_{n}^{\mathrm{nc}}} E[S(n) | \bb^n \in \cE_{n}^{\mathrm{nc}}] p(\bb^n \in \cE_{n}^{\mathrm{nc}}) \mathrm{d}\bb^n.
\label{eq:iterative6}
\end{align}

\noindent
(iii) Finally, observe that the second integral in \eqref{eq:iterative4} is the expected wealth growth of a TRP of length $n-i$, i.e.,
\begin{align}
E[S(n-i)] = \int_{\bb_{i+1}^n \in \cB_{n-i}(b,\eps)}  E[S_{i+1 \rightarrow n} | b(i) = b, \bb_{i+1}^n \in \cB_{n-i}(b,\eps)] p(\bb_{i+1}^n \in \cB_{n-i}(b,\eps)) \mathrm{d}\bb_{i+1}^n,
\label{eq:iterative7}
\end{align}
where $p(\bb_{i+1}^n \in \cB_{n-i}(b,\eps))=1$ by the definition of the set $\cB_{n-i}(b,\eps)$. 

Hence, if we apply \eqref{eq:iterative5}, \eqref{eq:iterative6} and \eqref{eq:iterative7} to \eqref{eq:iterative4}, we can write \eqref{eq:iterative-gain} as
\begin{align}
E[S(n)] = \sum_{i=1}^n P(\cE_{i}^{\mathrm{fc}}) R(i) E[S({n-i})] + P(\cE_{n}^{\mathrm{nc}}) T(n),
\label{eq:iterative8}
\end{align}
hence the proof concludes.
\end{proof}

Theorem~\ref{thm:iterative_wealth} provides a recursion to
iteratively calculate the expected wealth growth $E[S(n)]$, when
$R(\tau)$ and $T(\tau)$ are explicitly calculated for a TRP with
$(b-\eps, b+\eps)$. Hence, if we can obtain
$P\left(\cE_{\tau}^{\mathrm{fc}}\right) R(\tau)$ and
$P\left(\cE_{\tau}^{\mathrm{nc}}\right)T(\tau)$ for any $\tau$, then
\eqref{eq:iterative-gain} yields a simple iteration that provides the
expected wealth growth for any period $n$. We next give the explicit
definitions of the events $\cE_{\tau}^{\mathrm{fc}}$ and
$\cE_{\tau}^{\mathrm{nc}}$ in order to calculate the conditional
expectations $R(\tau)$ and $T(\tau)$. Following these definitions, we
calculate $P\left(\cE_{\tau}^{\mathrm{fc}}\right) R(\tau)$ and $
P\left(\cE_{\tau}^{\mathrm{nc}}\right) T(\tau)$ to evaluate
the expected wealth growth $E[S(\tau)]$, iteratively from
Theorem~\ref{thm:iterative_wealth} and find the the optimal TRP, i.e.,
optimal $b$ and $\eps$, by using a brute force search.
 
In the next section, we provide the explicit definitions for
$\cE_{\tau}^{\mathrm{fc}}$ and $\cE_{\tau}^{\mathrm{nc}}$, and define
the conditions for staying in the no rebalancing region or hitting
one of the boundaries to find the corresponding probabilities of these
events.

\subsection{Explicit Calculations of $R(n)$ and $T(n)$ \label{sec:explicit_calc}}
In this section, we first define the conditions for the market portfolios to
cross the corresponding thresholds and calculate the probabilities for the events
$\cE_{\tau}^{\mathrm{fc}}$ and $\cE_{\tau}^{\mathrm{nc}}$. We then
calculate the conditional expectations $R(n)$ and $T(n)$ as certain
multivariate Gaussian integrals. The explicit calculation of
multivariate Gaussian integrals are given in
Section~\ref{sec:multivariate-gaussian}.

To get the explicit definitions of the events $\cE_{\tau}^{\mathrm{fc}}$
and $\cE_{\tau}^{\mathrm{nc}}$, we note that we have two different
boundary hitting scenarios for a TRP, i.e., starting from the initial
portfolio $b$, the portfolio can hit $b-\eps$ or $b+\eps$. From $b$,
the portfolio crosses $b-\eps$ boundary if \be \frac{ b
  \prod_{t=1}^{\tau} (x_1(t)) }{ b \prod_{t=1}^{\tau} (x_1(t)) + (1-b)
  \prod_{t=1}^{\tau} (x_2(t)) } \leq b-\eps, \label{eq:cond1}
\end{equation}
where $\tau$ is the first time the crossing happens without ever
hitting any of the boundaries before.  Since $x_1(i) , x_2(i) > 0$
for all $i$, \eqref{eq:cond1} happens if \be \prod_{t=1}^\tau
\frac{x_2(t)}{x_1(t)} \geq
\frac{b(1-b+\eps)}{(1-b)(b-\eps)}, \label{eq:cond1_1} \ee which is
equivalent to
\[
\Pi_2(\tau) \geq \gamma_1 \Pi_1(\tau),
\]
where $\Pi_1(i) \defi \prod_{t=1}^{i} x_1(t)$, $\Pi_2(i) \defi
\prod_{t=1}^{i} x_2(t)$ and $\gamma_1 \defi \frac{b(1-b+\eps)}{(1-b)(b-\eps)}$. 
Since $\bx(i)$'s have log-normal distributions,
i.e., $\bx(t) \sim \ln {\cal N}(\bmu,\bsig^2)$,
$\Pi_1(i)$ and $\Pi_2(i)$ are log-normal, too
\cite{woods}. Furthermore, to calculate the required probabilities, we have
\begin{align}
p\left(\Pi_1(i) , \Pi_1(k-1), \Pi_1(k) \right) &= p\left(\Pi_1(i), \Pi_1(k-1) \right) p\left(\Pi_1(k)| \Pi_1(k-1), \Pi_1(i) \right) \nonumber \\
& = p\left( \Pi_1(i) \right) p\left(\Pi_1(k-1) | \Pi_1(i) \right) p\left(\Pi_1(k-1) x_1(k) | \Pi_1(k-1), \Pi_1(i)\right) \nonumber \\
&= p\left( \Pi_1(i) \right) p\left(\Pi_1(k-1) | \Pi_1(i) \right) p\left(\Pi_1(k) | \Pi_1(k-1) \right), \label{eq:markov_chain2}
\end{align}
$\forall i \in \{ 0, 1, \ldots, k-2\}$, where \eqref{eq:markov_chain2}
follows since $x(k)$ is independent of $\Pi_1(i)$ for $k > i$. Hence
$\Pi_1(i)$'s form a Markov chain such that $ \Pi_1(i) \leftrightarrow
\Pi_1(k-1) \leftrightarrow \Pi_1(k) \,\,\, \forall i \in \{ 0, 1,
\ldots, k-2\}$. Following the similar, steps we also obtain that $
\Pi_2(i) \leftrightarrow \Pi_2(k-1) \leftrightarrow \Pi_2(k) \,\,\,,
\forall i \in \{ 0, 1, \ldots, k-2\}$. We point out that by extending the 
definitions $\Pi_1$ and $\Pi_2$ one can obtain $\Pi_1, \Pi_2, \ldots, \Pi_m$ 
for the case $m>2$. Furthermore, taking the
logarithm of both sides of \eqref{eq:cond1_1} we have
\[
\Sigma_1^{\tau} \defi \sum_{t=1}^\tau z(t) \geq \theta_1,
\]
where $z(t) \defi \ln \left(\frac{x_2(t)}{x_1(t)}\right)$ and
$\theta_1 \defi \ln \frac{b(1-b+\eps)}{(1-b)(b-\eps)} = \ln
\gamma_1$. The partial sums of $z(t)$'s are defined as 
$\Sigma_i^k = \sum_{t=i}^k z(t)$ for notational simplicity. Since $\bx(t)
\sim \ln {\cal N}(\bmu,\bsig^2)$, $z(t)$'s are
Gaussian, i.e., $z(t) \sim {\cal N}(\mu,\sigma^2)$, where $\mu = \mu_2 - \mu_1$ 
and $\sigma^2 = \sigma_1^2 + \sigma_2^2$,
their sums, $\Sigma_i^k$'s, are
Gaussian too. Furthermore note that, $ \Sigma_1^k = \sum_{t=1}^k z(t) = \sum_{t=1}^k \ln \left(\frac{x_2(t)}{x_1(t)}\right) = \ln \left( \prod_{t=1}^k \frac{x_2(t)}{x_1(t)}\right) = \ln \frac {\Pi_2(k)}{\Pi_1(k)}$.

Similarly with an initial value $b$, market portfolio crosses $b+\eps$
boundary if \be \frac{ b \prod_{t=1}^{\tau} (x_1(t)) }{ b
  \prod_{t=1}^{\tau} (x_1(t)) + (1-b) \prod_{t=1}^{\tau} (x_2(t)) }
\geq b+\eps, \label{eq:cond2} \ee where $\tau$ is the first crossing
time without ever hitting any of the boundaries before.  Again, since
$x_1(i) , x_2(i) > 0$ for all $i$, \eqref{eq:cond2} happens if \be
\prod_{t=1}^\tau \frac{x_2(t)}{x_1(t)} \leq
\frac{b(1-b-\eps)}{(1-b)(b+\eps)}, \label{eq:cond2_1} \ee which can be
written of the form
\[
\Pi_2(t) \leq \gamma_2 \Pi_1(t).
\]
Equation \eqref{eq:cond2_1} yields
\[
\Sigma_1^{\tau} = \sum_{t=1}^\tau z(t) \leq \theta_2,
\]
where $\theta_2 \defi \ln \frac{b(1-b-\eps)}{(1-b)(b+\eps)} = \ln \gamma_2$.

Hence, we can explicitly describe the event that the market threshold
portfolio $(b-\eps,b+\eps)$ does not hit any of the thresholds for $\tau$
consecutive periods, $\cE_{\tau}^{\mathrm{nc}}$, as the intersection
of the events as
\begin{align}
\cE_{\tau}^{\mathrm{nc}} \defi \bigcap_{i=1}^{\tau} \, \{\Sigma_1^i \in [\theta_2, \theta_1]\} = \bigcap_{i=1}^{\tau} \, \{ \gamma_2 \Pi_1(i) \leq \Pi_2(i) \leq \gamma_1 \Pi_1(i)\}.
\label{E_nc_explicit}
\end{align}
Similarly, the event of the market threshold portfolio $(b-\eps,b+\eps)$ hitting any of the thresholds first time at the $\tau$-th period, $\cE_{\tau}^{\mathrm{fc}}$, can be defined as the intersections of the events
\begin{align}
\cE_{\tau}^{\mathrm{fc}} &\defi \bigcap_{i=1}^{\tau-1} \, \{\Sigma_1^i \in [\theta_2, \theta_1]\} \, \bigcap \, {\Big [}\{\Sigma_{\tau} \in [-\infty, \theta_2)\} \, \bigcup \, \{\Sigma_{\tau} \in (\theta_1,\infty]\} {\Big ]} \nonumber \\
&= \bigcap_{i=1}^{\tau-1} \, \{\gamma_2 \Pi_1(i) \leq \Pi_2(i) \leq \gamma_1 \Pi_1(i)\} \, \bigcap \, {\Big [}\{\Pi_2(\tau) \geq \gamma_1 \Pi_1(\tau)\} \, \bigcup \, \{\Pi_2(\tau) \leq \gamma_2 \Pi_1(\tau)\} {\Big ]},
\label{E_fc_explicit}
\end{align}
yielding the explicit definitions of the events $\cE_{\tau}^{\mathrm{fc}}$ 
in \eqref{E_fc_explicit} and $\cE_{\tau}^{\mathrm{nc}}$ in \eqref{E_nc_explicit}. 
The definitions of $\cE_{\tau}^{\mathrm{nc}}$ and $\cE_{\tau}^{\mathrm{fc}}$ 
can be readily extended for the case $m>2$ by using the updated definitions of 
$\Pi_1, \Pi_2, \ldots, \Pi_m$.

Since we have the quantitative definitions of the events
$\cE_{\tau}^{\mathrm{fc}}$ and $\cE_{\tau}^{\mathrm{nc}}$, we can
express the expected overall gain of portfolios with no hitting over
$\tau$-period, $T(\tau)$, as
\begin{align}
T(\tau) &= E{\Big [} S(\tau)  \,\, {\Big |} \,\, \cE_{\tau}^{\mathrm{nc}} {\Big ]} \nonumber \\
& =  E{\Big [}  b \prod_{t=1}^{\tau} [x_1(t)] + (1-b) \prod_{t=1}^{\tau} [x_2(t)] \,\, {\Big |} \,\, \cE_{\tau}^{\mathrm{nc}} {\Big ]} \nonumber \\
& =  E{\Big [} b \Pi_1(\tau) + (1-b) \Pi_2(\tau) \,\, {\Big |} \,\,  \cE_{\tau}^{\mathrm{nc}} {\Big ]}.
\label{T(n)-1}
\end{align}
The expectation $E{\Big [} b \Pi_1(\tau) + (1-b) \Pi_2(\tau) \,\, {\Big |} \,\,  \cE_{\tau}^{\mathrm{nc}} {\Big ]}$ 
can be expressed in an integral form as
\begin{align}
E{\Big [} b \Pi_1(\tau) + (1-b) \Pi_2(\tau) \,\, {\Big |} \,\,  \cE_{\tau}^{\mathrm{nc}} {\Big ]} = & \int_{0}^{\infty} \int_{0}^{\infty} \left( b \pi_1+ (1-b) \pi_2 \right) \,\, \nonumber \\
& \times P \left( \Pi_1(\tau) = \pi_1, \Pi_2(\tau) = \pi_2 \,\, {\Big |} \,\,  \cE_{\tau}^{\mathrm{nc}} \right) \,\mathrm{d}\pi_2\mathrm{d}\pi_1 \label{T(n)-2}
\end{align}
by the definition of conditional expectation. To extend this for the
case $m>2$, the double integral in the definition of $T_\tau$
\eqref{T(n)-2} is replaced by an $m$-dimensional integral over updated
random variables $\Pi_1, \Pi_2, \ldots, \Pi_m$. Combining
\eqref{T(n)-2} and \eqref{T(n)-1} yields
\begin{align}
T(\tau) &=  \int_{0}^{\infty} \int_{0}^{\infty} \left( b \pi_1+ (1-b) \pi_2 \right)
\,\, P \left( \Pi_1(\tau) = \pi_1, \Pi_2(\tau) = \pi_2 \,\, {\Big |} \,\,  \cE_{\tau}^{\mathrm{nc}} \right) \,\mathrm{d}\pi_2\mathrm{d}\pi_1 \nonumber\\
& = \frac{1}{P\left(\cE_{\tau}^{\mathrm{nc}}\right)} \int_{0}^{\infty} \int_{0}^{\infty} \left( b \pi_1+ (1-b) \pi_2 \right)
\,\, P \left( \Pi_1(\tau) = \pi_1, \Pi_2(\tau) = \pi_2 \right) \nonumber \\
& \times P \left( \cE_{\tau}^{\mathrm{nc}} \,\, {\Big |} \,\, \Pi_1(\tau) = \pi_1, \Pi_2(\tau) = \pi_2 \right) \,\mathrm{d}\pi_2\mathrm{d}\pi_1
\end{align}
by Bayes' theorem that $P(A|B)\,=\,\frac{P(B|A)P(A)}{P(B)}$. If we write 
the explicit definition of $\cE_{\tau}^{\mathrm{nc}}$ given in 
\eqref{E_nc_explicit}, then we obtain
\begin{align}
P\left(\cE_{\tau}^{\mathrm{nc}}\right) T(\tau) & = \int_{0}^{\infty} \int_{0}^{\infty} \left( b \pi_1+ (1-b) \pi_2 \right)
\,\, P \left( \Pi_1(\tau)=\pi_1, \Pi_2(\tau) = \pi_2 \right) P{\Big [} \gamma_2 \Pi_1(1) \leq \Pi_2(1) \leq \gamma_1 \Pi_1(1) \nonumber \\ &, \ldots , \gamma_2 \Pi_1(\tau) \leq \Pi_2(\tau) \leq \gamma_1 \Pi_1(\tau) {\Big |} \,\, \Pi_1(\tau)=\pi_1, \Pi_2(\tau) = \pi_2 {\Big ]} \,\mathrm{d}\pi_2\mathrm{d}\pi_1 \nonumber \\
& = \int_{0}^{\infty} \int_{\gamma_2 \pi_1}^{\gamma_1 \pi_1} \left( b \pi_1+ (1-b) \pi_2 \right)
\,\, P ( \Pi_1(\tau)=\pi_1, \Pi_2(\tau) = \pi_2) \nonumber\\
& \times P{\Big [} \gamma_2 \frac{\pi_1}{\prod_{t=2}^{\tau} x_1(t)} \leq \frac{\pi_2}{\prod_{t=2}^{\tau} x_2(t)} \leq \gamma_1 \frac{\pi_1}{\prod_{t=2}^{\tau} x_1(t)} , \gamma_2 \frac{\pi_1}{\prod_{t=3}^{\tau} x_1(t)} \leq \frac{\pi_2}{\prod_{t=3}^{\tau} x_2(t)} \leq \gamma_1 \frac{\pi_1}{\prod_{t=3}^{\tau} x_1(t)}\nonumber\\&, \ldots ,
\gamma_2 \frac{\pi_1}{x_1(\tau)} \leq \frac{\pi_2}{x_2(\tau)} \leq \gamma_1 \frac{\pi_1}{x_1(\tau)} {\Big ]} \,\mathrm{d}\pi_2\mathrm{d}\pi_1
\label{T(n)-3}
\end{align}
where \eqref{T(n)-3} follows by the definitions of $\Pi_1(i)$ and 
$\Pi_2(i)$, i.e., $\Pi_1(i) = \prod_{t=1}^{i} x_1(t)  = \frac{\Pi_1(\tau)}{\prod_{t=i+1}^{\tau} x_1(t)}$ and $\Pi_2(i) = \prod_{t=1}^{i} x_2(t)  = \frac{\Pi_2(\tau)}{\prod_{t=i+1}^{\tau} x_2(t)}$. If we 
rearrange the inequalities in \eqref{T(n)-3} to put the product 
terms together, which does not affect the direction of the 
inequality since all terms are positive, then we obtain
\begin{align}
& P\left(\cE_{\tau}^{\mathrm{nc}}\right)  T(\tau) = \int_{0}^{\infty} \int_{\gamma_2 \pi_1}^{\gamma_1 \pi_1} \left( b \pi_1+ (1-b) \pi_2 \right)
\,\, P ( \Pi_1(\tau)=\pi_1, \Pi_2(\tau) = \pi_2) P  {\Big [  } \frac{\pi_2}{\pi_1 \gamma_1} \leq \prod_{t=2}^{\tau} \frac{ x_2(t)}{x_1(t)} \leq \frac{\pi_2}{\pi_1 \gamma_2} ,\nonumber \\
&\frac{\pi_2}{\pi_1 \gamma_1} \leq \prod_{t=3}^{\tau} \frac{ x_2(t)}{x_1(t)} \leq \frac{\pi_2}{\pi_1 \gamma_2}, \ldots ,
\frac{\pi_2}{\pi_1 \gamma_1} \leq \frac{ x_2(\tau)}{x_1(\tau)} \leq \frac{\pi_2}{\pi_1 \gamma_2} {\Big ]} \,\mathrm{d}\pi_2\mathrm{d}\pi_1 \nonumber \\
& = \int_{0}^{\infty} \int_{\gamma_2 \pi_1}^{\gamma_1 \pi_1} \left( b \pi_1+ (1-b) \pi_2 \right)
\,\, P ( \Pi_1(\tau)=\pi_1, \Pi_2(\tau) = \pi_2) P  {\Big (  } \Sigma_2^{\tau} \in [\kappa -\theta_1, \kappa - \theta_2] , \Sigma_3^{\tau} \in [\kappa -\theta_1, \kappa -\theta_2], \nonumber \\
&\ldots, \Sigma_{\tau}^{\tau} \in [\kappa -\theta_1, \kappa -\theta_2] {\Big )} \,\mathrm{d}\pi_2\mathrm{d}\pi_1 \label{T(n)},
\end{align}
which follows from the definition of $\Sigma_i^{k}$ where $\kappa
\defi \ln \frac{\pi_2}{\pi_1}$. The first probability in \eqref{T(n)}
can be calculated as 
\begin{align}
	P \left( \Pi_1(\tau)=\pi_1, \Pi_2(\tau) = \pi_2\right) &= P \left( \Pi_1(\tau)=\pi_1\right) P\left( \Pi_2(\tau) = \pi_2\right) \nonumber \\
	&=\frac{1}{\pi_1 \sqrt{2\pi \, \tau \sigma_1^2}} e^{-\frac{(\ln \pi_1 - \tau \mu_1)^2}{2 \,\tau \sigma_1^2}} +
	  \frac{1}{\pi_1 \sqrt{2 \pi \,\tau \sigma_2^2}} e^{-\frac{(\ln \pi_2 - \tau \mu_2)^2}{2 \,\tau \sigma_2^2}} \label{eq:log-norm-pdf}
\end{align}
which follows since $\Pi_1(\tau) \defi \prod_{t=1}^{\tau} x_1(t)$ and $\Pi_2(\tau) \defi \prod_{t=1}^{\tau} x_2(t)$ and
we have $\Pi_1(\tau) \sim \ln {\cal N}(\tau \mu_1,\tau \sigma_1^2)$ and $\Pi_2(\tau) \sim \ln {\cal N}(\tau \mu_2,\tau \sigma_2^2)$.
The corresponding terms in
\eqref{T(n)} are written as a multi variable integral calculated in
Section~\ref{sec:multivariate-gaussian}.

Following similar steps, we can obtain the expected overall gain 
$R(\tau)$ as
\begin{align}
R(\tau) & = E\left[ S(\tau)  \,\, {\Big |} \,\, \cE_{\tau}^{\mathrm{fc}} \right] \nonumber \\
& =  E\left[  b \prod_{t=1}^{\tau} [x_1(t)] + (1-b) \prod_{t=1}^{\tau} [x_2(t)] - 2c(b - b^2) | \prod_{t=1}^{\tau} [x_1(t)] - \prod_{t=1}^{\tau} [x_2(t)] | \,\, {\Big |} \,\, \cE_{\tau}^{\mathrm{fc}} \right].
\label{R(n)-1}
\end{align}
The conditional expectation $E\left[ S(\tau) \,\, {\Big |} \,\,  \cE_{\tau}^{\mathrm{fc}} \right]$ 
can also be expressed in an integral form as
\begin{align}
E\left[ S(\tau) \,\, {\Big |} \,\,  \cE_{\tau}^{\mathrm{fc}} \right] = & \int_{0}^{\infty} \int_{0}^{\infty} S(\tau) \,\, P \left( \Pi_1(\tau) = \pi_1, \Pi_2(\tau)  = \pi_2 \,\, {\Big |} \,\,  \cE_{\tau}^{\mathrm{fc}} \right) \,\,\mathrm{d}\pi_2\mathrm{d}\pi_1, \label{R(n)-4}
\end{align}
which follows from the definition of conditional expectation. 
Combining \eqref{R(n)-4} and \eqref{R(n)-1} yields
\begin{align}
R(\tau) & =  \int_{0}^{\infty} \int_{0}^{\infty} S(\tau) \,\, P \left( \Pi_1(\tau) = \pi_1, \Pi_2(\tau) = \pi_2 \,\, {\Big |} \,\,  \cE_{\tau}^{\mathrm{fc}} \right) \,\mathrm{d}\pi_2\mathrm{d}\pi_1 \nonumber\\
& = \frac{1}{P\left(\cE_{\tau}^{\mathrm{fc}}\right)} \int_{0}^{\infty} \int_{0}^{\infty} S(\tau) \,\, P \left( \Pi_1(\tau) = \pi_1, \Pi_2(\tau) = \pi_2 \right) \nonumber \\
& \times P \left( \cE_{\tau}^{\mathrm{fc}} \,\, {\Big |} \,\, \Pi_1(\tau) = \pi_1, \Pi_2(\tau) = \pi_2 \right) \,\mathrm{d}\pi_2\mathrm{d}\pi_1,
\label{R(n)-2}
\end{align}
where \eqref{R(n)-2} follows from the Bayes' theorem. Note that the 
definition of $R(\tau)$ \eqref{R(n)-2} can be extended for the case 
$m>2$ by replacing the double integral with an $m$-dimensional integral 
over the updated random variables $\Pi_1, \Pi_2, \ldots, \Pi_m$. If we 
replace the event $\cE_{\tau}^{\mathrm{fc}}$ with its explicit definition 
in \eqref{E_fc_explicit}, then we get
\begin{align}
&P\left(\cE_{\tau}^{\mathrm{fc}}\right) R(\tau)  = \int_{0}^{\infty} \int_{0}^{\infty} \left( \zeta_1 \pi_1+ \zeta_2 \pi_2 \right)
\,\, P \left( \Pi_1(\tau)=\pi_1, \Pi_2(\tau) = \pi_2 \right) P  {\Big [} \gamma_2 \Pi_1(1) \leq \Pi_2(1) \leq \gamma_1 \Pi_1(1), \ldots , \nonumber \\
&\gamma_2 \Pi_1(\tau-1) \leq \Pi_2(\tau-1) \leq \gamma_1 \Pi_1(\tau-1), \gamma_1 \Pi_1(\tau) \leq \Pi_2(\tau) \,\, {\Big |} \,\, \Pi_1(\tau)=\pi_1, \Pi_2(\tau) = \pi_2 {\Big ]} \,\mathrm{d}\pi_2\mathrm{d}\pi_1 \nonumber \\
+ & \int_{0}^{\infty} \int_{0}^{\infty} \left( \zeta_3 \pi_1+ \zeta_4 \pi_2 \right) P \left( \Pi_1(\tau)=\pi_1, \Pi_2(\tau) = \pi_2 \right) P  {\Big [} \gamma_2 \Pi_1(1) \leq \Pi_2(1) \leq \gamma_1 \Pi_1(1), \ldots , \nonumber \\
&\gamma_2 \Pi_1(\tau-1) \leq \Pi_2(\tau-1) \leq \gamma_1 \Pi_1(\tau-1), \gamma_2 \Pi_1(\tau) \geq \Pi_2(\tau) \,\, {\Big |} \,\, \Pi_1(\tau)=\pi_1, \Pi_2(\tau) = \pi_2 {\Big ]} \,\mathrm{d}\pi_2\mathrm{d}\pi_1, \label{R(n)-3}
\end{align}
where $\zeta_1 \defi b - 2 c (b - b^2)$, $\zeta_2 = 1- b + 2 c (b - b^2)$ , $\zeta_3 = b + 2 c (b - b^2)$ and $\zeta_4 = 1- b - 2 c (b - b^2)$. We next calculate the first
integral in \eqref{R(n)-3} and the second integral follows similarly. 

By the definitions of $\Pi_1(i)$ and $\Pi_2(i)$, we have $\Pi_1(i) = \prod_{t=1}^{i} x_1(t)  = \frac{\Pi_1(\tau)}{\prod_{t=i+1}^{\tau} x_1(t)}$ and $\Pi_2(i) = \prod_{t=1}^{i} x_2(t)  = \frac{\Pi_2(\tau)}{\prod_{t=i+1}^{\tau} x_2(t)}$, 
hence the first integral in \eqref{R(n)-3} can be written as
\begin{align}
&\int_{0}^{\infty} \int_{\gamma_1 \pi_1}^{\infty} \left( \zeta_1 \pi_1+ \zeta_2 \pi_2 \right)  P ( \Pi_1(\tau)=\pi_1, \Pi_2(\tau) = \pi_2) P  {\Big [} \gamma_2 \frac{\pi_1}{\prod_{t=2}^{\tau} x_1(t)} \leq \frac{\pi_2}{\prod_{t=2}^{\tau} x_2(t)} \leq \gamma_1 \frac{\pi_1}{\prod_{t=2}^{\tau} x_1(t)},  \nonumber \\ 
&\gamma_2 \frac{\pi_1}{\prod_{t=3}^{\tau} x_1(t)} \leq \frac{\pi_2}{\prod_{t=3}^{\tau} x_2(t)} \leq \gamma_1 \frac{\pi_1}{\prod_{t=3}^{\tau} x_1(t)}, \ldots ,\gamma_2 \frac{\pi_1}{x_1(\tau)} \leq \frac{\pi_2}{x_2(\tau)} \leq \gamma_1 \frac{\pi_1}{x_1(\tau)} {\Big ]} \,\mathrm{d}\pi_2\mathrm{d}\pi_1.
\label{R(n)-5}
\end{align}
If we gather the product terms in \eqref{R(n)-5} into the same fraction, 
then we obtain
\begin{align}
& \int_{0}^{\infty} \int_{\gamma_1 \pi_1}^{\infty}  \left( \zeta_1 \pi_1+ \zeta_2 \pi_2 \right)
\,\, P ( \Pi_1(\tau)=\pi_1, \Pi_2(\tau) = \pi_2) P  {\Big [  } \frac{\pi_2}{\pi_1 \gamma_1} \leq \prod_{t=2}^{\tau} \frac{ x_2(t)}{x_1(t)} \leq \frac{\pi_2}{\pi_1 \gamma_2} ,\nonumber \\
&\frac{\pi_2}{\pi_1 \gamma_1} \leq \prod_{t=3}^{\tau} \frac{ x_2(t)}{x_1(t)} \leq \frac{\pi_2}{\pi_1 \gamma_2}, \ldots ,
\frac{\pi_2}{\pi_1 \gamma_1} \leq \frac{ x_2(\tau)}{x_1(\tau)} \leq \frac{\pi_2}{\pi_1 \gamma_2} {\Big ]} \,\mathrm{d}\pi_2\mathrm{d}\pi_1 \\
& = \int_{0}^{\infty} \int_{\gamma_1 \pi_1}^{\infty}  \left( \zeta_1 \pi_1+ \zeta_2 \pi_2 \right)
\,\, P ( \Pi_1(\tau)=\pi_1, \Pi_2(\tau) = \pi_2) P  {\Big (  } \Sigma_2^{\tau} \in [\kappa -\theta_1, \kappa - \theta_2] , \Sigma_3^{\tau} \in [\kappa -\theta_1, \kappa -\theta_2], \nonumber \\
&\ldots, \Sigma_{\tau}^{\tau} \in [\kappa -\theta_1, \kappa -\theta_2] {\Big )} \,\mathrm{d}\pi_2\mathrm{d}\pi_1, \label{R(n)-6}
\end{align}
which follows from the definition of $\Sigma_i^{k}$ where $\kappa \defi \ln \frac{\pi_2}{\pi_1}$. Following similar steps 
that yields \eqref{R(n)-6}, we can calculate \eqref{R(n)-3} as
\begin{align}
P\left(\cE_{\tau}^{\mathrm{fc}}\right) R(\tau) = & \int_{0}^{\infty} \int_{\gamma_1 \pi_1}^{\infty}  \left( \zeta_1 \pi_1+ \zeta_2 \pi_2 \right)
	\,\, P ( \Pi_1(\tau)=\pi_1, \Pi_2(\tau) = \pi_2) P  {\Big (  } \Sigma_2^{\tau} \in [\kappa -\theta_1, \kappa - \theta_2] , \nonumber \\
&\Sigma_3^{\tau} \in [\kappa -\theta_1, \kappa -\theta_2], \ldots, \Sigma_{\tau}^{\tau} \in [\kappa -\theta_1, \kappa -\theta_2] {\Big )} \,\mathrm{d}\pi_2\mathrm{d}\pi_1 \nonumber \\
& + \int_{0}^{\infty} \int_{0}^{\gamma_2 \pi_1} \left( \zeta_3 \pi_1+ \zeta_4 \pi_2 \right)
\,\, P ( \Pi_1(\tau)=\pi_1, \Pi_2(\tau) = \pi_2) P  {\Big (  } \Sigma_2^{\tau} \in [\kappa -\theta_1, \kappa - \theta_2] ,  \nonumber \\
&\Sigma_3^{\tau} \in [\kappa -\theta_1, \kappa -\theta_2], \ldots, \Sigma_{\tau}^{\tau} \in [\kappa -\theta_1, \kappa -\theta_2] {\Big )} \,\mathrm{d}\pi_2\mathrm{d}\pi_1, \label{R(n)}
\end{align}
where the probability $P ( \Pi_1(\tau)=\pi_1, \Pi_2(\tau) = \pi_2)$ can 
be obtained via \eqref{eq:log-norm-pdf}. Hence to calculate 
$P\left(\cE_{\tau}^{\mathrm{nc}}\right) T(\tau)$ and $P\left(\cE_{\tau}^{\mathrm{fc}}\right) R(\tau)$, 
we need to calculate the probability $P  {\Big (  } \Sigma_2^{\tau} \in [\kappa -\theta_1, \kappa - \theta_2] , \Sigma_3^{\tau} \in [\kappa -\theta_1, \kappa -\theta_2], \ldots, \Sigma_{\tau}^{\tau} \in [\kappa -\theta_1, \kappa -\theta_2] {\Big )}$ 
in \eqref{T(n)} and \eqref{R(n)}. 

Following from the definition of $\Sigma_{i}^k$s, we have
\begin{align}
p(\Sigma_{i}^k , \Sigma_{i+1}^k, \Sigma_{j}^k ) &= p(\Sigma_{i+1}^k, \Sigma_{j}^k ) p(\Sigma_{i}^k | \Sigma_{i+1}^k, \Sigma_{j}^k ) \nonumber \\
& = p( \Sigma_{j}^k ) p(\Sigma_{i+1}^k | \Sigma_{j}^k ) p(\Sigma_{i+1}^k + z(i)| \Sigma_{i+1}^k, \Sigma_{j}^k ) \nonumber \\
&= p( \Sigma_{j}^k ) p(\Sigma_{i+1}^k | \Sigma_{j}^k ) p(\Sigma_{i}^k | \Sigma_{i+1}^k) \label{eq:markov_chain}
\end{align}
$\forall i \in \{ 0, 1, \ldots, k-2\}$, where \eqref{eq:markov_chain} follows 
since $z(i)$ is independent of $\Sigma_{j}^{k}$ for $j > i$. Then, $\Sigma_i^k$'s 
form a Markov chain such that $\Sigma_j^k \leftrightarrow \Sigma_{i+1}^k \leftrightarrow \Sigma_{i}^k \,\,\, \forall i \in \{ 0, 1, \ldots, k-2\}$ 
and $j>i$.
Hence, we can write the probability
\begin{align}
& P  {\Big (  } \Sigma_2^{\tau} \in [\kappa -\theta_1, \kappa - \theta_2] , \Sigma_3^{\tau} \in [\kappa -\theta_1, \kappa -\theta_2], \ldots, \Sigma_{\tau}^{\tau} \in [\kappa -\theta_1, \kappa -\theta_2] {\Big )} \nonumber \\
& = \int_{\kappa -\theta_1}^{ \kappa - \theta_2}  \int_{\kappa -\theta_1}^{ \kappa - \theta_2} \ldots  \int_{\kappa -\theta_1}^{ \kappa - \theta_2} P(\Sigma_{\tau}^{\tau} = s_{1}, \Sigma_{\tau-1}^{\tau} = s_{2}, \ldots, \Sigma_{2}^{\tau} = s_{\tau-1}) \,\mathrm{d}s_{\tau-1}\mathrm{d}s_{\tau-2}\ldots\mathrm{d}s_1 \nonumber \\
& =  \int_{\kappa -\theta_1}^{ \kappa - \theta_2}  \int_{\kappa -\theta_1}^{ \kappa - \theta_2} \ldots  \int_{\kappa -\theta_1}^{ \kappa - \theta_2} P(\Sigma_{2}^{\tau} = s_{\tau-1} | \Sigma_{3}^{\tau} = s_{\tau-2}) P(\Sigma_{3}^{\tau} = s_{\tau-2} | \Sigma_{4}^{\tau} = s_{\tau-3}) \ldots \nonumber\\
& P(\Sigma_{\tau-1}^{\tau} = s_{2} | \Sigma_{\tau}^{\tau} = s_{1})P( \Sigma_{\tau}^{\tau} = s_{1}) \,\mathrm{d}s_{\tau-1}\mathrm{d}s_{\tau-3}\ldots\mathrm{d}s_2\mathrm{d}s_1, \label{trprob1}
\end{align}
where \eqref{trprob1} follows by the chain rule  and $\Sigma_i$'s form a 
Markov chain. We can express the conditional probabilities in \eqref{trprob1}, 
which are of the form $P(\Sigma_{i}^{\tau} = s_{\tau-i} | \Sigma_{i+1}^{\tau} = s_{\tau-i-1})$, as
\begin{align}
P(\Sigma_{i}^{\tau} = s_{\tau-i+1} | \Sigma_{i+1}^{\tau} = s_{\tau-i}) & = P(\Sigma_{i+1}^{\tau} + z(i) = s_{\tau-i+1} | \Sigma_{i+1}^{\tau} = s_{\tau-i}) \nonumber \\
& = P(s_{\tau-i} + z(i) = s_{\tau-i+1} | \Sigma_{i+1}^{\tau} = s_{\tau-i}) \nonumber \\
& = P(z(i) = s_{\tau-i+1} - s_{\tau-i} | \Sigma_{i+1}^{\tau} = s_{\tau-i}) \nonumber \\
& = P(z(i) = s_{\tau-i+1} - s_{\tau-i}) \label{trprob2}
\end{align}
where \eqref{trprob2} follows from the independence of $z(i)$ and $z(k)$'s 
for $i < k \leq \tau$ or the independence of $z(i)$ and 
$\Sigma_{i+1}^{\tau} = \sum_{k=i+1}^{\tau} z(k)$. If we replace 
\eqref{trprob2} with the conditional probabilities in \eqref{trprob1} 
and use $P( \Sigma_{\tau}^{\tau} = s_{1}) = P( z(\tau)= s_{1})$, then we obtain
\begin{align}
& P  {\Big (  } \Sigma_2^{\tau} \in [\kappa -\theta_1, \kappa - \theta_2] , \Sigma_3^{\tau} \in [\kappa -\theta_1, \kappa -\theta_2], \ldots, \Sigma_{\tau}^{\tau} \in [\kappa -\theta_1, \kappa -\theta_2] {\Big )} \nonumber \\
& = \int_{\kappa -\theta_1}^{ \kappa - \theta_2}  \int_{\kappa -\theta_1}^{ \kappa - \theta_2} \ldots  \int_{\kappa -\theta_1}^{ \kappa - \theta_2} f_z(s_{\tau-1} - s_{\tau-2})f_z(s_{\tau-2} - s_{\tau-3}) \ldots f_z(s_2 - s_1)f_z(s_1)  \,\mathrm{d}s_{\tau-1}\mathrm{d}s_{\tau-2}\ldots\mathrm{d}s_2\mathrm{d}s_1 \nonumber \\
& = \int_{\kappa -\theta_1}^{ \kappa - \theta_2}  \int_{\kappa -\theta_1}^{ \kappa - \theta_2} \ldots  \int_{\kappa -\theta_1}^{ \kappa - \theta_2} ({\frac{1}{2 \pi \sigma^2}})^{\frac{\tau-1}{2}} e^{\frac{-1}{2 \sigma^2}\sum_{i=2}^{\tau-1} (s_i - s_{i-1} - \mu)^2 + (s_1 - \mu)^2} \,\mathrm{d}s_{\tau-1}\mathrm{d}s_{\tau-2}\ldots\mathrm{d}s_2\mathrm{d}s_1, \label{eq:ndim-gauss}
\end{align}
where \eqref{eq:ndim-gauss} follows since $z(i)$'s are Gaussian, 
$z \sim {\cal N}(\mu,\sigma^2)$, i.e., $f_z(.)$ is the normal 
distribution. Hence in order to iteratively calculate the expected 
wealth growth of a TRP, we need to calculate the multivariate Gaussian 
integral given in \eqref{eq:ndim-gauss}, which is investigated in the 
next section.

\subsection{Multivariate Gaussian Integrals\label{sec:multivariate-gaussian}}
In order to complete calculation of the iterative equation in
\eqref{eq:iterative-gain}, we next evaluate the definite multivariate
Gaussian integral given in \eqref{eq:ndim-gauss} on the
multidimensional $[\kappa -\theta_1 , \kappa - \theta_2]^n$ space. We
emphasize that the corresponding multivariate integral cannot be
calculated using common diagonalizing methods\cite{Timm}. Although,
in \eqref{eq:ndim-gauss}, the coefficient matrix of the multivariate
integral is symmetric positive-definite, common diagonalizing methods
cannot be directly applied since the integral bounds after a
straightforward change of variables depend on $y_i$. However,
\eqref{eq:ndim-gauss} can be represented as certain error functions of
Gaussian distributions.

We note that the multivariate Gaussian integral given in
\eqref{eq:ndim-gauss} is the ``non-central multivariate normal integral'' or
non-central MVN integral \cite{Genz2009} and general MVN integrals 
are in the form \cite{Genz2009}
\begin{equation}
	\Phi_k(\ba,\bb,\bSig) = \frac {1}{\sqrt{|\bSig| (2\pi)^k}} \int_{a_1}^{b_1}  \int_{a_2}^{b_2} \ldots \int_{a_k}^{b_k} e^{\frac{-1}{2} {\bx}^T {\bSig}^{-1} \bx} \,\mathrm{d}x_{k}\ldots\mathrm{d}x_2\mathrm{d}x_1, \label{mvn_integral}
\end{equation}
where $\bSig$ is a symmetric, positive definite covariance matrix. In our
case, \eqref{eq:ndim-gauss} is a non-central MVN integral which can be written
in the form \eqref{mvn_integral}, where $k = \tau -1$ and the inverse of the 
covariance matrix is given by
\begin{figure}
{\scriptsize
\begin{tabular}[t]{|l|}
\hline
\textbf{A Pseudo-code of QMC Algorithm for MVN Integrals:}\\
\hline
{\bf 1.} get $\bSig$, $\ba$, $\bb$, $N$, $M$ and $\alpha$  \\
{\bf 2.} compute lower triangular Cholesky factor $L$ for $\bSig$, permuting $\ba$ and $\bb$, 
and rows and columns of $\bSig$ for variable prioritization.  \\
{\bf 3.} initialize $P = 0$, $N = 0$, $V = 0$, and $q = \sqrt{p}$ with $\bp = (2,3,5,\ldots,p_k)$ where $p_j$ is the $j$-th prime.  \\
{\bf 4.} for $i=1,2,\ldots,M$ do \\
\hspace{0.3in} $I_i = 0$ and generate uniform random $\bDelta \in [0,1]^k$ shift vector.\\
\hspace{0.3in} for $j=1,2,\ldots,N$ do\\
\hspace{0.6in} $\bw = |2(j\bq + \bDelta) - \bone|$ ,  \\
\hspace{0.6in} $d_1 = \Phi \left(\frac{a_1}{l_{1,1}}\right)$ , $e_1 = \Phi \left(\frac{b_1}{l_{1,1}}\right)$ and $f_1 = e_1 - d_1$. \\
\hspace{0.6in} for $m = 2, 3,\ldots, k$ do\\
\hspace{0.9in} $y_{m-1} = \Phi^{-1} (d_{m-1} + w_{m-1}(e_{m-1} - d_{m-1})) $, \\
\hspace{0.9in} $d_{m} = \Phi \left( \frac{a_m - \sum_{n=1}^{m-1} l_{m,n} y_j} { l_{m,m} } \right)$, \\
\hspace{0.9in} $e_{m} = \Phi \left( \frac{b_m - \sum_{n=1}^{m-1} l_{m,n} y_j} { l_{m,m} } \right)$, \\
\hspace{0.9in} $f_{m} = (e_m - d_m) f_{m-1}$. \\
\hspace{0.6in} endfor\\
\hspace{0.6in} $I_i = I_i + (f_m - I_i)/j.$ \\
\hspace{0.3in} endfor\\
\hspace{0.3in} $\sigma = (I_i - t)/i$, $P = P + \sigma$, $V = (i-2)V/i + \sigma^2$ and $E = \alpha \sqrt{V}$  \\
endfor\\
{\bf 5.} output $P \approx \Phi_k(\ba,\bb,\bSig)$ with error estimate $E$. \\
\hline
\end{tabular}}
\caption{A randomized QMC algorithm proposed in \cite{Genz2009} to compute MVN probabilities for hyper-rectangular regions.}
\label{fig:randomize_alg}
\end{figure}
\begin{align*}
{\bSig}^{-1} = \begin{bmatrix}2 & -1 &  \\ -1 & 2 & -1 \\ & \ddots & \ddots & \ddots \\ & & -1 & 2 & -1\\ & & &-1& 2  \end{bmatrix}
\end{align*}
which is a symmetric positive definite matrix with $|\bSig| = 1$, the lower bound vector 
is of the form, $\ba= [a_1,\ldots,a_{\tau-1}]^T$,
\begin{align*}
\ba = \begin{bmatrix} \kappa - \theta_1 - \mu  \\ \kappa - \theta_1 - 2 \mu \\ \vdots \\ \kappa - \theta_1 - (\tau-1)\mu  \end{bmatrix}
\end{align*}
and the upper bound vector is given by,  $\bb= [b_1,\ldots,b_{\tau-1}]^T$,
\begin{align*}
\bb = \begin{bmatrix} \kappa - \theta_2 - \mu  \\ \kappa - \theta_2 - 2 \mu \\ \vdots \\ \kappa - \theta_2 - (\tau-1)\mu  \end{bmatrix}
\end{align*}
where $-k \mu$ terms in the lower and the upper bounds follow from the 
non-central property of \eqref{eq:ndim-gauss}. We emphasize that the MVN 
integral in \eqref{mvn_integral} cannot be calculated 
in a closed form \cite{Genz2009} and most of the
results on this integral correspond to either special cases or coarse
approximations \cite{Blake73,Genz2009}. Hence, in this paper, we use
the randomized QMC algorithm, provided in
Fig.~\ref{fig:randomize_alg} \cite{Genz2009} for completeness, to
compute MVN probabilities over hyper rectangular regions. Here, the
algorithm uses a periodization and randomized QMC rule \cite{Richtmyer},
where the output error estimate $E$ in Fig.~\ref{fig:randomize_alg} is
the usual Monte Carlo standard error based on $N$ samples of the
randomly shifted QMC rule, and scaled by the confidence factor
$\alpha$. We observe in our simulations that the algorithm in
Fig.~\ref{fig:randomize_alg} produce satisfactory results on the
historical data \cite{Kozat11}. We emphasize that different algorithms
can be used instead of the Quasi-Monte Carlo (QMC) algorithm to
calculate the multivariable integrals in \eqref{eq:ndim-gauss},
however, the derivations still hold.

\section{Maximum-Likelihood Estimation of Parameters of the Log-Normal Distribution \label{sec:mle_est}} 

In this section, we give the MLEs for the mean and variance of the
log-normal distribution using the sequence of price relative vectors,
which are used sequentially in the Simulations section to evaluate the
optimal TRPs. Since the investor observes the sequence of price
relatives sequentially, he or she needs to estimate $\bmu$ and $\bsig$
at each investment period to find the maximizing $b$ and $\eps$.
Without loss of generality we provide the MLE for $x_1(t)$, where the
MLE for $x_2(t)$ directly follows.

For these derivations, we assume that we observed a sequence of price
relative vectors of length $N$, i.e., $(x_1(1),x_1(2), \ldots$,
$x_1(N))$. Note that the sample data need not to belong to $N$
consecutive periods such that the sequential representation is chosen
for ease of presentation. Then, we find the parameters $\mu_1$ and
$\sigma_1^2$ that maximize the log-likelihood function
\begin{align*}
\ln \cL (\mu_1,\sigma_1^2 \, | \, x_1(1), x_1 (2), \ldots, x_1(N)) = \ln f(x_1(1), x_1 (2), \ldots, x_1(N) \, | \, \mu_1,\sigma_1^2 ) = \sum_{i=1}^N \ln f(x_1(i)\,| \,\mu_1,\sigma_1),
\end{align*}
where $f(x|\mu_1,\sigma_1^2) = \frac{1}{x \sqrt{2 \pi {\sigma_1}^2}}
e^{-\frac{(\ln x - \mu_1)^2}{2 {\sigma_1}^2}}$. The log-likelihood
function in \eqref{eq:mle_1} can also be written as
\begin{align}
\ln \cL (\mu_1,\sigma_1^2 \, | \, x_1(1), x_1 (2), \ldots, x_1(N)) &= \sum_{i=1}^N \ln \frac{1}{x_1(i) \sqrt{2 \pi {\sigma_1}^2}} e^{-\frac{(\ln x_1(i) - \mu_1)^2}{2 {\sigma_1}^2}} \nonumber \\
& = \sum_{i=1}^N \ln \frac{1}{x_1(i) \sqrt{2 \pi {\sigma_1}^2}} - \sum_{i=1}^N \frac{(\ln x_1(i) - \mu_1)^2}{2 {\sigma_1}^2}. \label{eq:mle_1}
\end{align}
We start with maximizing the log-likelihood function $\ln \cL$ with
respect to $\mu_1$, i.e., find the estimator $\hat{\mu_1}$ that
satisfies $\frac{\partial \ln \cL }{\partial \mu_1} = 0$. If we take the
partial derivative of the expression in \eqref{eq:mle_1} with respect
to $\mu_1$, then we obtain
\begin{align*}
\frac{\partial \ln \cL}{\partial \mu_1} = \sum_{i=1}^N \frac{\ln x_1(i) - \mu_1}{{\sigma_1}^2}.
\end{align*}
Hence  $\mu_1$, which satisfies $\frac{\partial \cL}{\partial \mu_1} = 0$, or the ML estimator $\hat{\mu_1}$ of $\mu_1$, can be found as
\begin{align}
\hat{\mu_1} = \frac{1}{N} \sum_{i=1}^N {\ln x_1(i)}.
\label{eq:mle_mu}
\end{align}
To find the ML estimator of the variance $\sigma_1^2$, we find
$\hat{\sigma_1^2}$ that satisfies $\frac{\partial \ln \cL }{\partial
  \sigma_1^2} = 0$.  Since $\mu_1$ that satisfies $\frac{\partial
  \hat{l}}{\partial \mu_1} = 0$ in \eqref{eq:mle_mu} does not depend
on $\sigma_1^2$, we can use it in \eqref{eq:mle_1}. Let us define
$\bar{x_1} = \sum_{i=1}^N \frac{\ln x_1(i)}{N}$ for
notational clarity. By replacing $\bar{x_1}$ with $\mu_1$ in
\eqref{eq:mle_mu} and taking the partial derivative of the expression
with respect to $\sigma_1^2$, we obtain
\begin{align*}
\frac{\partial \ln \cL}{\partial \sigma_1^2} = -\frac{N}{2\sigma_1^2} + \frac{1}{2(\sigma_1^2)^2} \sum_{i=1}^N (\ln x_1(i) - \bar{x_1})^2.
\end{align*}
Hence 
\begin{align}
\hat{\sigma_1^2} = \frac{1}{N} \sum_{i=1}^N (\ln x_1(i) - \bar{x_1})^2.
\label{eq:mle_sigma}
\end{align}
Following similar steps, the ML estimators for $x_2(t)$ yield
\begin{align}
\hat{\mu_2} = \frac{1}{N} \sum_{i=1}^N \ln x_2(i),
\label{eq:mle_mu_2}
\end{align}
and
\begin{align}
\hat{\sigma_2^2} = \frac{1}{N} \sum_{i=1}^N (\ln x_2(i) - \bar{x_2})^2,
\label{eq:mle_sigma_2}
\end{align}
where $\bar{x_2} \defi \sum_{i=1}^N \frac{\ln x_2(i)}{N}$. Note that the ML estimators $\hat{\mu_1}$, $\hat{\sigma_1^2}$,
$\hat{\mu_2}$ and $\hat{\sigma_2^2}$ are consistent \cite{Ruppert},
i.e., they converge to the true values as the size of the data set goes to
infinity, i.e., $N \rightarrow \infty$ \cite{woods}.

\section{Simulations \label{sec:sim}}
In this section, we illustrate the performance our algorithm under
different scenarios. We first use TRPs over simulated data of two
stocks, where each stock is generated from a log-normal distribution.
We then continue to test the performance over the historical ``Ford -
MEI Corporation'' stock pair chosen for its volatility \cite{helmbold} from the New
York Stock Exchange.  As the final set of experiments, we use our
algorithm over the historical data set from \cite{cover91} and
illustrate the average performance. In all these trials, we compare
the performance of our algorithm with portfolio selection strategies
from \cite{cover91, iyengar_uni, Kozat11}.
\begin{figure}
  \centering
  \subfloat[]{\label{fig:fakec025}\includegraphics[width=140mm,height=65mm]{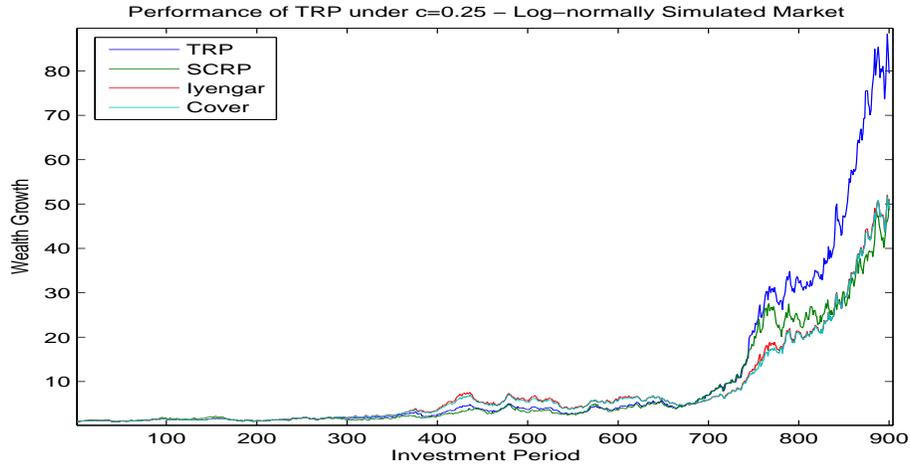}}                
  \\ \subfloat[]{\label{fig:fakec01}\includegraphics[width=140mm,height=65mm]{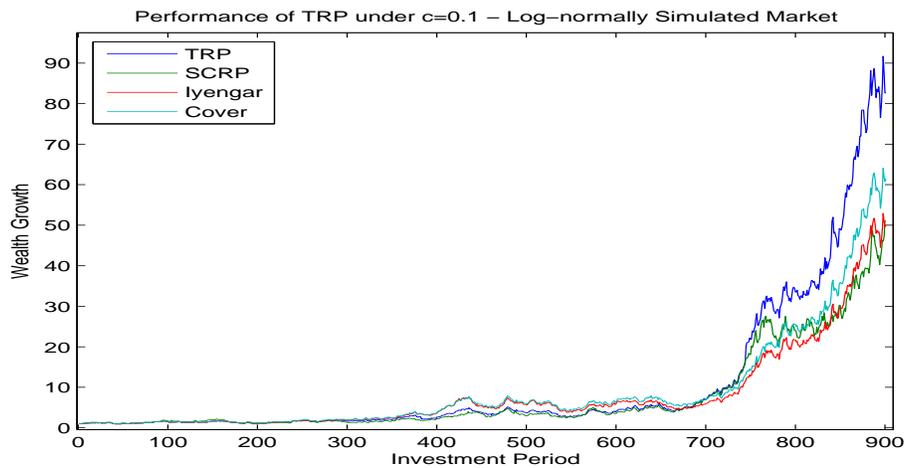}}
  \caption{\small Performance of various portfolio investment algorithms on a Log-normally simulated two-stock market. (a) Wealth growth under hefty transaction cost (c=0.025). (b) Wealth growth under moderate transaction cost (c=0.01).}
  \label{fig:fake}
\end{figure}

\begin{figure}
  \centering
  \subfloat[]{\label{fig:pairc025}\includegraphics[width=140mm,height=65mm]{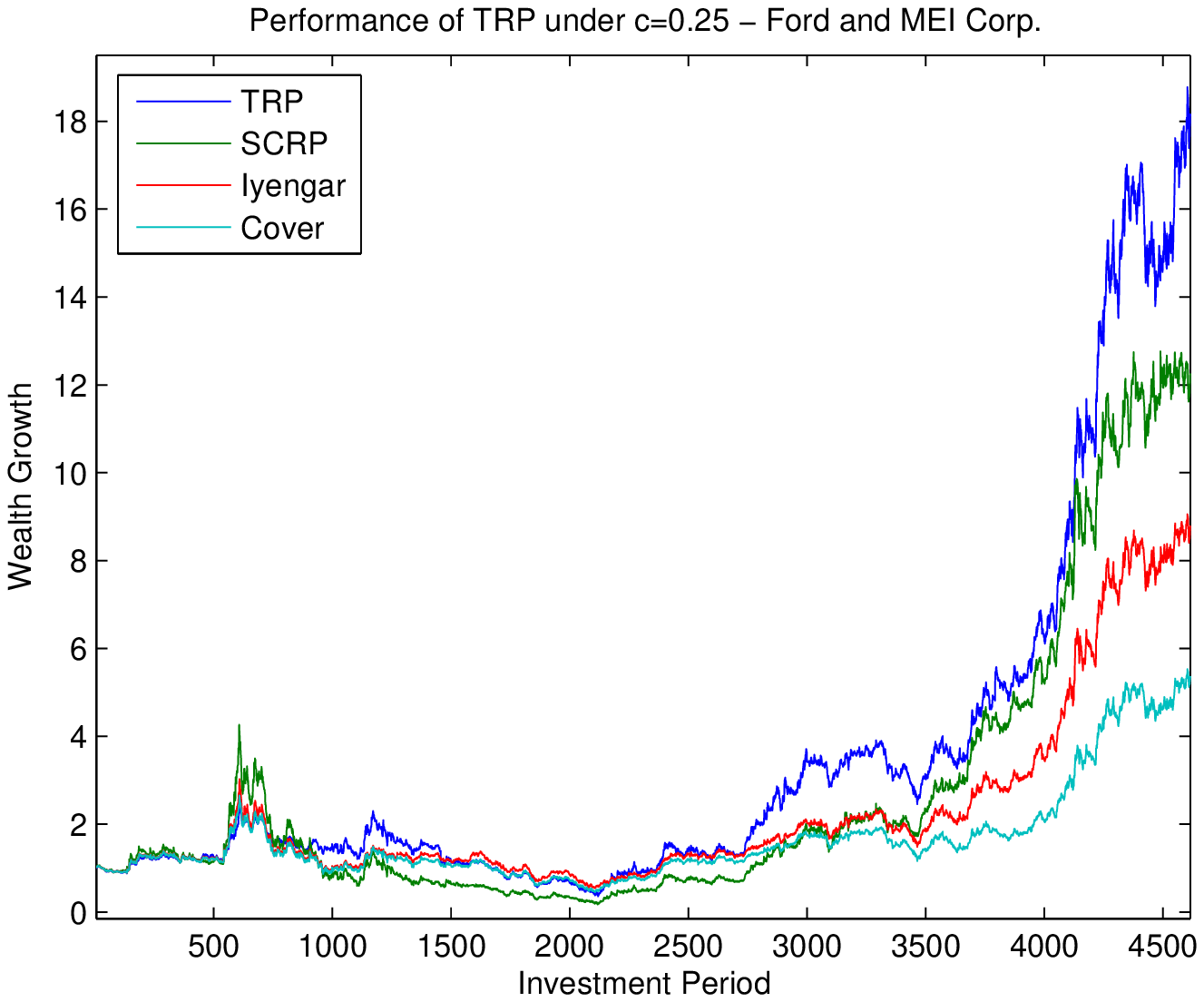}}                
  \\ \subfloat[]{\label{fig:pairc01}\includegraphics[width=140mm,height=65mm]{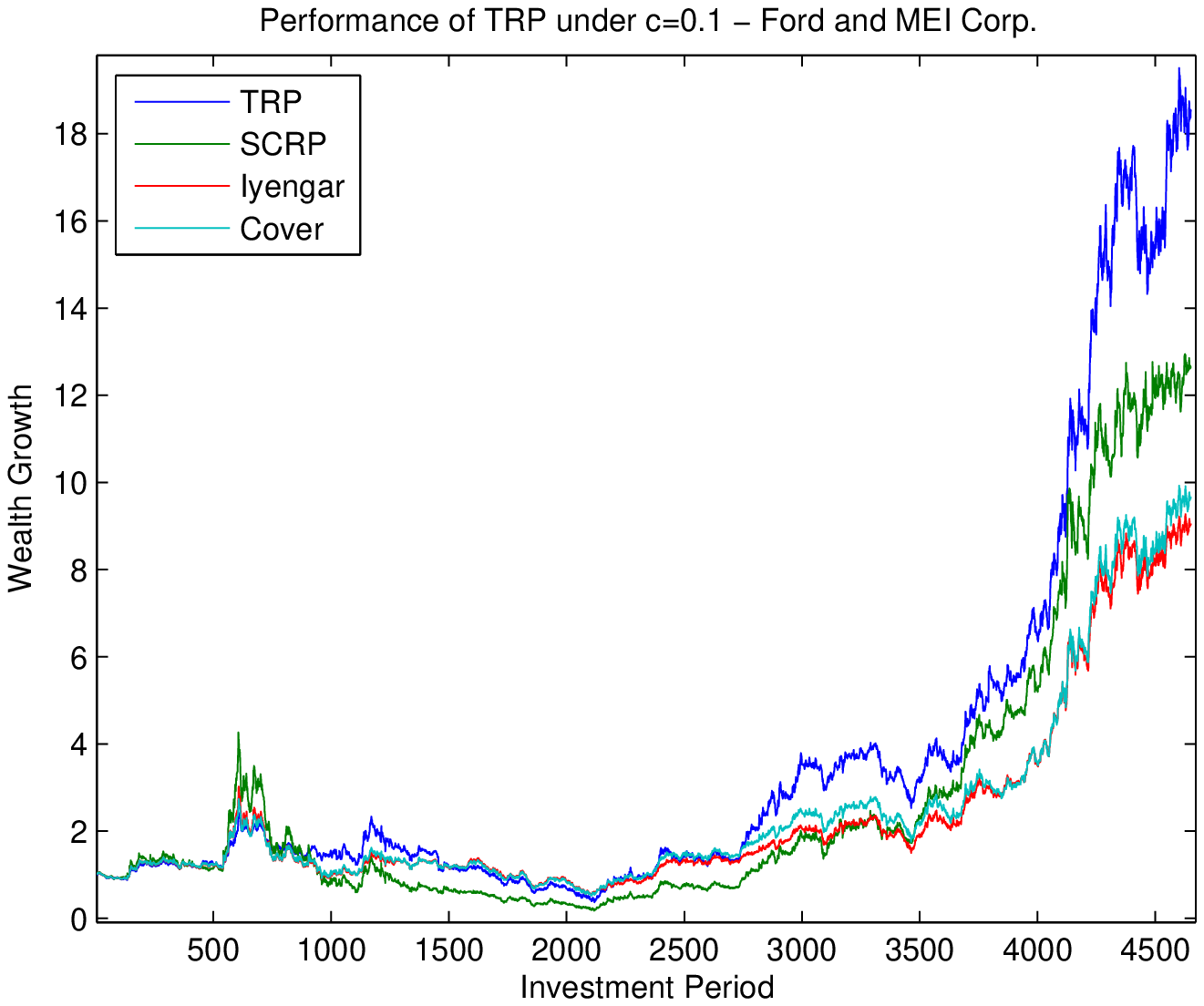}}
  \caption{\small Performance of various portfolio investment algorithms on Ford - MEI Corporation pair. (a) Wealth growth under hefty transaction cost (c=0.025). (b) Wealth growth under moderate transaction cost (c=0.01).}
  \label{fig:pair}
\end{figure}
In the first example, each stock is generated from a log-normal
distribution such that $x_1(t) \sim \ln {\cal N}(0.006,0.05)$ and
$x_2(t) \sim \ln {\cal N}(0.003,0.05)$, where the mean and variance
values are arbitrarily selected. We observe that the results do not
depend on a particular choice of model parameters as long as they
resemble real life markets. We simulate the performance over 1100
investment periods. Since the mean and variance parameters are not
known by the investor, we use the ML estimators from
Section~\ref{sec:mle_est}, which are then used to determine the target
portfolio $b$ and the threshold value $\eps$.  We start by calculating
the ML estimators using the initial 200 samples and find the target
portfolio $\bb=[b\;\; 1-b]^T$ and the threshold $\eps$ that maximize
the expected wealth growth by a brute-force search.  Then, we use the
corresponding $b$ and $\eps$ during the following 200 samples. In
similar lines, we calculate and use the optimal TRP for a total of 900
days, where $b$ and $\epsilon$ are estimated over every window of 200
samples and used in the following window of 200 samples. We choose a
window of size 200 samples to get reliable estimates for the means and
variances based on the size of the overall data.  In
Fig.~\ref{fig:fake}, we show the performances of: this sequential TRP
algorithm ``TRP'', the Cover's universal portfolio selection algorithm
\cite{cover91} ``Cover'', the Iyengar's universal portfolio algorithm
\cite{iyengar_uni} ``Iyengar'' and a semiconstant rebalanced portfolio
(SCRP) algorithm \cite{Kozat11} ``SCRP'', where the parameters are
chosen as suggested in \cite{Kozat11}. As seen in Fig.~\ref{fig:fake},
the TRP with the parameters sequentially calculated using the ML
estimators is the best rebalancing strategy among the others as
expected from our derivations.  In Fig.~\ref{fig:fakec025} and
Fig.~\ref{fig:fakec01}, we present results for a mild transaction cost
$c=0.01$ and a hefty transaction cost $c=0.025$, respectively, where
$c$ is the fraction paid in commission for each transaction, i.e.,
$c=0.01$ is a $1\%$ commission. We observe that the performance of the
TRP algorithm is better than the other algorithms for these
transaction costs. However, the relative gain is larger for the large
transaction cost since the TRP approach, with the optimal parameters
chosen as in this paper, can hedge more effectively against the
transaction costs.
\begin{figure}
  \centering
  \subfloat[]{\label{fig:avec025}\includegraphics[width=140mm,height=65mm]{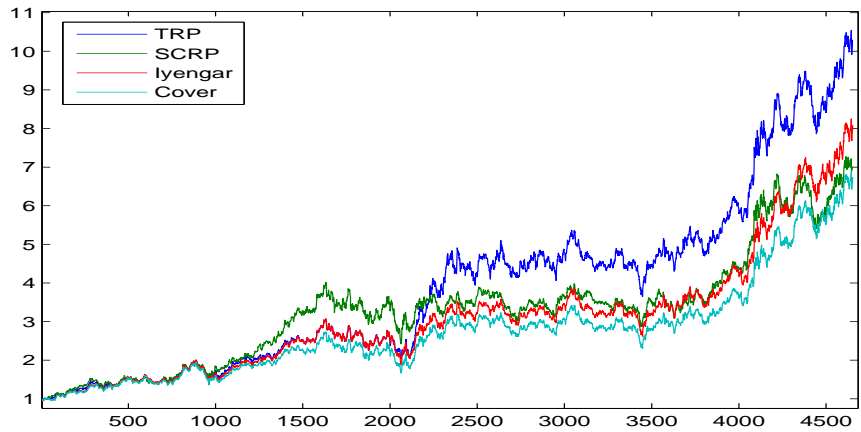}}                
  \\ \subfloat[]{\label{fig:avec01}\includegraphics[width=140mm,height=65mm]{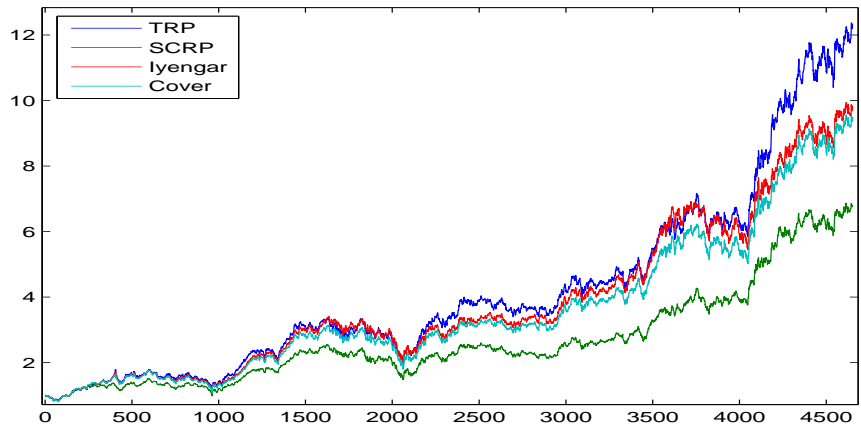}}
  \caption{\small Performance of various portfolio investment algorithms on Ford - MEI Corporation pair. (a) Wealth growth under hefty transaction cost (c=0.025). (b) Wealth growth under moderate transaction cost (c=0.01).}
  \label{fig:ave}
\end{figure}

As the next example, we apply our algorithm to historical data from
\cite{cover91} from the New York Stock Exchange collected over a
22-year period.  We first apply algorithms on the ``Ford - MEI
Corporation'' pair as shown in Fig.~\ref{fig:pair}, which are chosen
because of their volatility \cite{helmbold}. In Fig.~\ref{fig:pair},
we plot the wealth growth of: the sequential TRP algorithm with the
optimal parameters sequentially calculated, the Cover's universal
portfolio, the Iyengar's universal portfolio and the SCRP algorithm
with the suggested parameters in \cite{Kozat11}. We use the ML
estimators to choose the optimal TRP as in the first set of
experiments, however, since the historical data contains 5651 days we
use a window of size 1000 days.  Hence, the performance results are
shown over 4651 days. As seen from Fig.~\ref{fig:pair}, the proposed
TRP algorithm significantly outperforms other algorithms for this data
set. Similar to the simulated data case, we investigate the
performance of the TRP algorithm under different transaction costs,
i.e., a moderate transaction cost $c=0.01$ in Fig.~\ref{fig:pairc01}
and a hefty transaction cost $c=0.025$ in
Fig.~\ref{fig:pairc025}. Comparing the results from the
Fig.~\ref{fig:pairc025} and Fig.~\ref{fig:pairc01}, we conclude that
the TRP with the optimal sequential parameter selection can better
handle the transaction costs when the stocks are volatile for this
experiment.

Finally, to remove any bias on a particular stock pair, we show the
average performance of the TRP algorithm over randomly selected stock
pairs from the historical data set from \cite{cover91}. The total set
includes 34 different stocks, where the Iroquois stock is removed due
to its peculiar behavior. We first randomly select pairs of stocks and
invest using: the sequential TRP algorithm with the sequential ML
estimators, the Cover's universal portfolio algorithm, the Iyengar's
universal portfolio algorithm and the SCRP algorithm. The sequential
selection of the optimal TRP parameters are performed similar to the
previous case, i.e., we use ML estimators on an investment block of
1000 days and use the calculated optimal TRP in the next block of 1000
days. For each stock pair, we simulate the performance of the
algorithms over 4651 days.  In Fig.~\ref{fig:ave}, we present the
wealth achieved by these algorithms, where the results are averaged
over 10 independent trials.  We present the achieved wealth over
random sets of stock pairs under a moderate transaction cost $c=0.01$
in Fig.~\ref{fig:avec01} and a hefty transaction cost $c=0.025$ in
Fig.~\ref{fig:avec025}. As seen from the figures, the TRP algorithm
with the ML estimators readily outperforms the other strategies
under different transaction costs on this historical data set.

\section{Conclusion}
\label{sec:conclusion}
In this paper, we studied an important financial application, the
portfolio selection problem, from a signal processing perspective. We
investigated the portfolio selection problem in i.i.d. discrete time
markets having a finite number of assets, when the market levies
proportional transaction fees for both buying and selling stocks. We
introduced algorithms based on threshold rebalanced portfolios that
achieve the maximal growth rate when the sequence of price relatives
have the log-normal distribution from the well-known Black-Scholes
model \cite{investment}. Under this setup, we provide an iterative
relation that efficiently and recursively calculates the expected
wealth in any i.i.d. market over any investment period. The terms in
this recursion are evaluated by a certain multivariate Gaussian
integral. We then use a randomized algorithm to calculate the given
integral and obtain the expected growth. This expected growth is then
optimized by a brute force method to yield the optimal target
portfolio and the threshold to maximize the expected wealth over any
investment period. We also provide a maximum-likelihood estimator to
estimate the parameters of the log-normal distribution from the
sequence of price relative vectors. As predicted from our derivations,
we significantly improve the achieved wealth over portfolio selection
algorithms from the literature on the historical data set from \cite{cover91}.

\bibliographystyle{IEEEbib}

\bibliography{msaf_references}

\end{document}